\documentclass[twocolumn]{aastex63}
\usepackage{hyperref}
\usepackage[utf8]{inputenc}
\usepackage[encapsulated]{CJK}

\usepackage{newunicodechar}
\DeclareRobustCommand{\okina}{%
  \raisebox{\dimexpr\fontcharht\font`A-\height}{%
    \scalebox{0.8}{`}%
  }%
}
\newunicodechar{ʻ}{\okina}

\shorttitle{blast}
\shortauthors{}
\graphicspath{{./}{figures/}}

\begin{document}

\title{Blast: a Web Application for Characterizing the Host Galaxies of Astrophysical Transients }


\author[0000-0002-6230-0151]{D.~O.~Jones}
\altaffiliation{Joint first author contribution.}
\affiliation{Institute for Astronomy, University of Hawaiʻi, 640 N.~Aʻohoku Pl., Hilo, HI 96720, USA}

\author[0000-0002-1052-6749]{P.~McGill}
\altaffiliation{Joint first author contribution.}
\affiliation{Department of Astronomy and Astrophysics, University of California, Santa Cruz, CA 93105, USA}
\affiliation{Space Science Institute, Lawrence Livermore National Laboratory, 7000 East Ave., Livermore, CA 94550, USA}

\author[0000-0003-2545-9195]{T.\ A.\ Manning}
\affiliation{National Center for Supercomputing Applications, University of Illinois Urbana-Champaign, Urbana, IL 61801, USA}

\author[0000-0003-4906-8447]{A.\ Gagliano}
\affiliation{The NSF AI Institute for Artificial Intelligence and Fundamental Interactions}
\affiliation{Center for Astrophysics \textbar{} Harvard \& Smithsonian, Cambridge, MA 02138, USA}
\affiliation{Department of Physics, Massachusetts Institute of Technology, Cambridge, MA 02139, USA}

\author[0000-0001-9269-5046]{B.\ Wang (\begin{CJK*}{UTF8}{gbsn}王冰洁\ignorespacesafterend\end{CJK*})}
\affiliation{Department of Astronomy \& Astrophysics, The Pennsylvania State University, University Park, PA 16802, USA}
\affiliation{Institute for Computational \& Data Sciences, The Pennsylvania State University, University Park, PA 16802, USA}
\affiliation{Institute for Gravitation and the Cosmos, The Pennsylvania State University, University Park, PA 16802, USA}

\author[0000-0003-4263-2228]{D.~A.~Coulter}
\affiliation{Space Telescope Science Institute, Baltimore, MD 21218, USA}

\author[0000-0002-2445-5275]{R.~J.~Foley}
\affiliation{Department of Astronomy and Astrophysics, University of California, Santa Cruz, CA 93105, USA}

\author[0000-0001-6022-0484]{G.~Narayan}
\affiliation{Department of Astronomy, University of Illinois Urbana-Champaign, Urbana, IL 61801, USA}
\affiliation{Center for Astrophysical Surveys, National Center for Supercomputing Applications, Urbana, IL, 61801, USA}

\author[0000-0002-5723-8023]{V.~A.~Villar}
\affiliation{Center for Astrophysics \textbar{} Harvard \& Smithsonian, Cambridge, MA 02138, USA}
\affiliation{The NSF AI Institute for Artificial Intelligence and Fundamental Interactions}

%

%


\author{L.~Braff}
\affiliation{Department of Astronomy and Astrophysics, University of California, Santa Cruz, CA 93105, USA}

\author[0000-0003-2348-483X]{A.~W.~Engel}
\affiliation{Pacific Northwest National Laboratory, 902 Battelle Blvd, Richland, WA 99354, USA}

\author[0000-0002-0786-7307]{D.~Farias}
\affiliation{DARK, Niels Bohr Institute, University of Copenhagen, Jagtvej 128, DK-2200 Copenhagen {\O} Denmark}

\author[0000-0001-8483-9089]{Z.~Lai}
\affiliation{Department of Astronomy and Astrophysics, San Francisco State University, San Francisco, CA 94132}

\author{K.~Loertscher}
\affiliation{Department of Astronomy and Astrophysics, University of California, Santa Cruz, CA 93105, USA}

\author[0009-0004-7605-8484]{J.~Kutcka}
\affiliation{Department of Astronomy and Astrophysics, University of California, Santa Cruz, CA 93105, USA}

\author[0009-0005-6323-0457]{S.~Thorp} 
\affiliation{The Oskar Klein Centre, Department of Physics, Stockholm University, AlbaNova University Centre, SE 106 91 Stockholm, Sweden}

\author[0000-0003-1576-0830]{J.\ Vazquez}
\affiliation{Department of Astronomy, University of Illinois Urbana-Champaign, Urbana, IL 61801, USA}

\begin{abstract}

Characterizing the host galaxies of astrophysical transients is important to many areas of astrophysics, including constraining the progenitor systems of core-collapse supernovae, correcting Type Ia supernova distances, and probabilistically classifying transients without photometric or spectroscopic data.
Given the increasing transient discovery rate in the coming years, there is substantial utility in providing public, transparent, reproducible, and automatic characterization for large samples of transient host galaxies. Here we present {\tt Blast}, a web application that ingests live streams of transient alerts, matches transients to their host galaxies, and performs photometry on coincident archival imaging data of the host galaxy.  The photometry is then used to infer both global host-galaxy properties and galaxy properties within 2~kpc of the transient location by using the {\tt Prospector} Bayesian inference framework, with an acceleration in evaluation speed achieved via simulation-based inference. 
{\tt Blast} provides host-galaxy properties to users via a web browser or an application program interface. The software can be extended to support alternative photometric or SED-fitting algorithms, and can be scaled via an asynchronous worker queue across multiple compute nodes to handle the processing of large volumes of transient alerts for upcoming transient surveys.  {\tt Blast} has been ingesting newly discovered transients from the Transient Name Server since mid-2024, and has currently measured SED parameters for more than 6000 transients.  The service is publicly available at \url{https://blast.scimma.org/}.
\end{abstract}


\section{Introduction} \label{sec:intro}

The properties of transients are fundamentally linked with the characteristics of the galaxies in which their progenitor stars form and evolve \citep[e.g.,][]{Timmes03,Sullivan2010, Rigault2013, Anderson15, MorenoRaya16, Taggart2021, Qin24}. Therefore, characterizing and understanding the host galaxies of astrophysical transients can play a crucial role in understanding the transients themselves. Understanding the properties of transient host galaxies, both globally and near the location of the transient, can aid in understanding the progenitor systems of core collapse supernovae (SNe) \citep[e.g.,][]{Anderson12,Sanders2012,Anderson15,Hosseinzadeh2019,Taggart2021}, probe rapidly evolving objects \citep{Wiseman20b}, and mitigate bias in Type Ia SN (SN\,Ia) distance measurements \citep{Kelly2010, Lampeitl2010, Sullivan2010} that are then incorporated in measurements of cosmological parameters \citep[e.g.,][]{Scolnic18, Abbott2019, Jones2019, Brout22, Moller2024}. 
Taking into account the correlations between host galaxy and transient properties is also an important step in planning upcoming transient surveys, including those using the Vera C.\ Rubin Observatory and the Nancy G.\ Roman Space Telescope \citep{Lokken2023}.

Host-galaxy properties are also useful in classifying transients \citep[e.g.,][]{Foley2013,Kisley2023} because they provide discriminating class information beyond approaches that solely use transient photometry \citep[e.g.,][]{Boone2019, Qu2022}. Real-time classification of SNe in the hours to days after explosion can be particularly vital for making decisions about subsequent spectroscopic and photometric followup \citep[e.g.,][]{Kasliwal2019, Muthukrishna19, Aleo2022, Coulter2022, Coulter2023}, in order to probe the SN progenitor star \citep[e.g.,][]{Kilpatrick2021,Tinyanont2022}, or close-in circumstellar material \citep[e.g.,][]{Jacobson2022}.  At early times, when very little light curve data is available, the extra leverage gained from host information can be particularly important \citep{Gagliano23}.

Historically, optimally leveraging host-galaxy information has proven difficult for four main reasons. Firstly, obtaining the necessary sample sizes needed to fully understand host galaxies and transient correlations has been challenging, particularly for rare types  \citep[e.g., ][]{Perley2016b, Perley2016}. Secondly, biases in transient survey design have affected the measured host-galaxy demographics \citep[for example, surveys that target pre-selected bright galaxies, e.g.,][]{Leaman2011, Li2011}. Thirdly, photometric/spectroscopic coverage of the host galaxy has often been limited to optical wavelengths \citep[e.g.,][]{Jones2018}. Finally, heterogeneous, and sometimes inconsistent, methods and models used to fit broad-band photometry of the host galaxy can in some cases lead to poor and biased estimation of host galaxy parameters and their uncertainties \citep[][]{Leja2017,Leja2019,Lower2020,Hand22,Lower22,Nagaraj22,Wang23b}. 

The challenges relating to small transient samples and the design of transient surveys have been largely addressed with the recent transition from narrow-field galaxy-targeted surveys to wide-field time-domain surveys, including ASAS-SN \citep{Shappee14, Kochanek17}, ATLAS \citep{Tonry18}, Gaia Photometric Science Alerts \citep{Hodgkin2021}, the Panoramic Survey Telescope and Rapid Response (Pan-STARRS) Survey for Transients \citep[PSST;][]{Huber2015}, the Zwicky Transient Facility \citep[ZTF;][]{Bellm2019}, and the the Young Supernova Experiment \citep[YSE;][]{Jones2021}, among others. These modern transient surveys have delivered a larger, more homogeneous and exponentially growing transient discovery rate over recent years with 20,713 transients reported in 2023 alone.\footnote{\url{https://www.wis-tns.org/stats-maps}}  The discovery rate is expected to increase further still with the advent of the Vera Rubin Observatory's Legacy Survey of Space and Time \citep[LSST;][]{LSSTScienceCollaboration2009}. 

For all of these newly discovered transients, there exists coincident, readily-available archival imaging data. This multi-wavelength archival imaging data ranges from the infrared (2MASS; \citealt{Skrutskie2006}, WISE; \citealt{Wright2010}) through the optical (SDSS, \citealt{York2000,Blanton2017}; DESI Legacy Imaging Surveys, including the Dark Energy Camera Legacy Survey, \citealp{Dey2019}; Pan-STARRS, \citealt{Chambers16,Flewelling20,Magnier20}) and the ultraviolet (GALEX; \citealt{Martin2005}). These data provide the means to consistently and robustly measure the host galaxy spectral energy distribution \citep[SED; e.g.,][]{Muller-Bravo2022, Bradley2022,Duarte23,Meldorf23} and, when combined with Bayesian fitting methods \citep{Johnson2021, Speagle2020}, infer the properties of the host galaxies of transients as they are discovered. These available host-galaxy characterization tools and archival imaging data make addressing the final two challenges outlined above possible, at least in theory. In practice, the characterization of transient host galaxies requires significant compute when performing photometry on archival images and fitting models to the resulting host-galaxy SEDs.

In recent years the transient community has invested heavily in the development of transient alert brokers (e.g., MARS, \citealt{Brown2013}; ANTARES, \citealt{Saha2014}; AMPEL, \citealt{Nordin2019}; Fink, \citealt{Moller2021}; ALeRCE, \citealt{Forster2021}; Lasair; \citealt{Smith19}). These brokers provide a stream of alerts of transients to the community, along with some basic discovery information.  This stream of alerts can then be ingested by Target and Observation Manager (TOM) platforms (e.g., {\tt YSE-PZ}, \citealt{Coulter2022, Coulter2023}; the RoboNet Microlensing System, \citealt{Tsapras2009}; the PESSTO Marshall, \citealt{Smartt2015}; the Supernova Exchange, \citealt{Howell2017}; the TOM-toolkit, \citealt{Street2018}; the Global Relay of Observatories Watching Transients Happen Marshall, \citealt{Kasliwal2019}; the Transient Science Server, \citealt{Smith2020}; and the SNAD ZTF object viewer, \citealt{Malanchev2023}), where typically computationally inexpensive value-added services can be run on the transient (e.g., host-matching, \citealt{Gagliano2021}; light curve classification, \citealt{Aleo2022}) to track and manage followup decisions and data. However, none of the transient alert brokers of TOM platforms currently provide consistent and automatic host galaxy characterization.

In this paper,
we present 
{\tt Blast}: an open-source web application that can ingest live streams of transient alerts and automatically measure the physical properties of its host galaxy stellar population globally and locally within 2~kpc of the transient location, using available archival imaging data.
In Section \ref{sec:blast} we describe the system architecture of {\tt Blast}, user interface, and host galaxy characterization pipeline. In Section \ref{sec:verification} we validate {\tt Blast}'s pipeline against previous studies. In Section \ref{sec:discussion} we present brief statistics for {\tt Blast}'s current set of transient events and discuss future challenges for transient host galaxy characterization in the era of upcoming large transient surveys.  In Section \ref{sec:conclusions} we conclude.

\section{Blast}\label{sec:blast}

{\tt Blast}\footnote{Available at \url{https://blast.scimma.org/} with documentation at \url{https://blast.readthedocs.io/}.} is a web application that ingests, stores, and processes transients, transient metadata, and transient host-galaxy data. This data is available to users through a front end web application and via an Application Program Interface (API). In this section, we will first cover {\tt Blast}'s underlying data model and architecture (Section \ref{sec:architecture}). Then we will detail {\tt Blast}'s tasks processing system and the specific processes that {\tt Blast} uses to characterize transient host galaxies (Section \ref{sec:processing}).  Finally, we will describe the details of the web application (Section \ref{sec:app}) and its deployment (Section \ref{sec:deployment}).

\begin{figure*}
    \centering
    \includegraphics[width=7in]{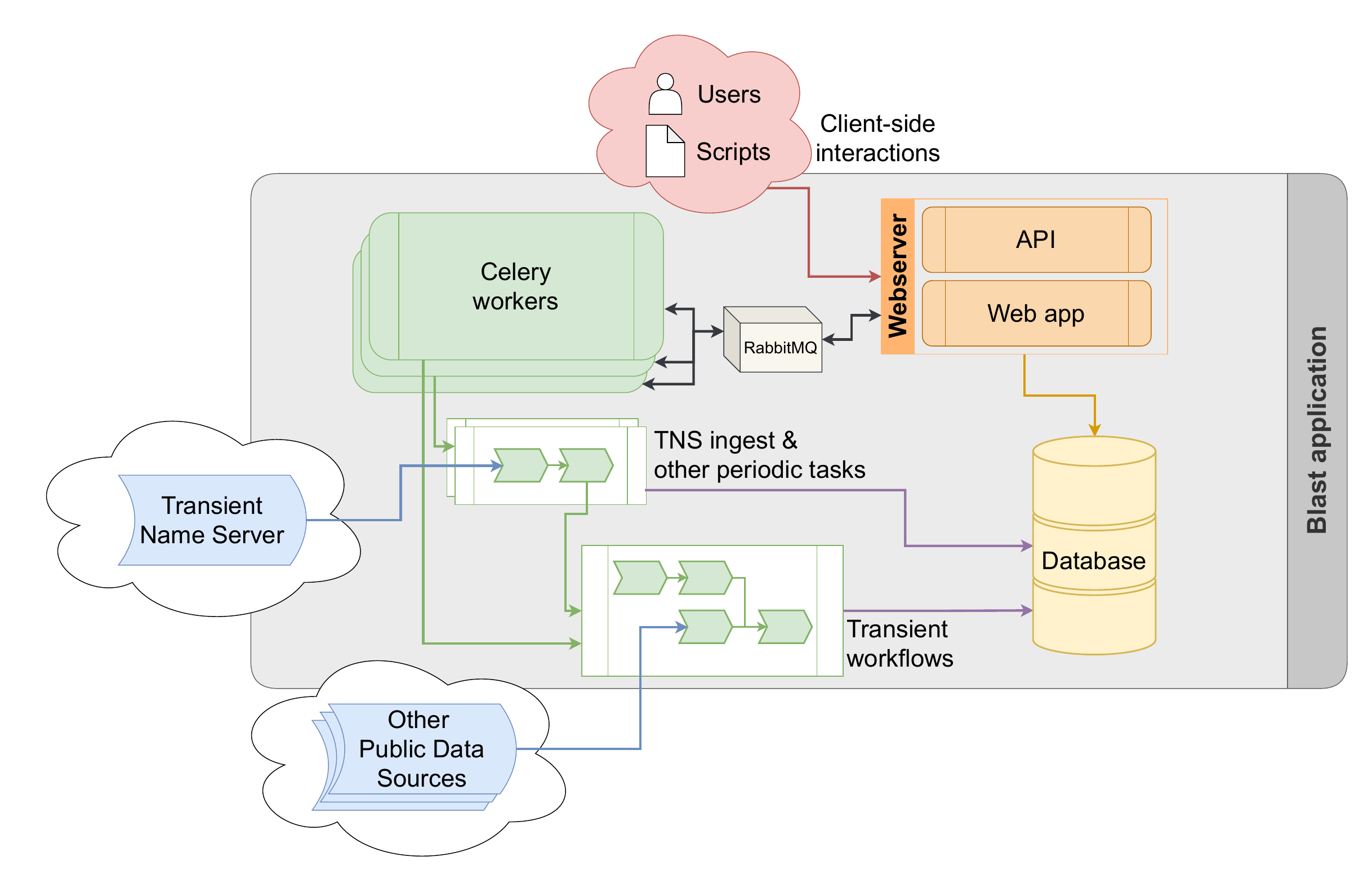}
    \caption{Schematic of the {\tt Blast} system architecture. Arrows indicate the direction of data flow or interaction. Celery workers periodically execute tasks including one that polls the Transient Name Server API and stores new transient alert data in the database. Each alert triggers a transient processing workflow that executes as a set of Celery tasks across all available workers. The workflow tasks download required data from public online resources as needed and perform the necessary computation tasks. Users can view and download the resulting data via the web application or programmatically via the API.}
    \label{fig:system}
\end{figure*}

\subsection{Software Architecture}\label{sec:architecture}

{\tt Blast} is built using the Django framework \citep{django19:django} and leverages its extensive ecosystem of available apps and integrations, including Celery and the Django REST Framework \citep{djangorestframework}. The {\tt Blast} application is comprised of several components: (1) the Django webserver that runs both the frontend web app and the backend API server, (2) a database for primary data storage, (3) a pool of Celery workers, which are processes that execute asynchronous data processing tasks, (4) a Celery Beat instance for periodic task scheduling, and (5) a RabbitMQ message broker that provides a communication mechanism between the web app server and the Celery workers via the Advanced Message Queuing Protocol (AMQP).  The overview of the {\tt Blast} system architecture is shown in Figure \ref{fig:system}. 

\begin{figure}
    \centering
    \includegraphics[width=3.4in]{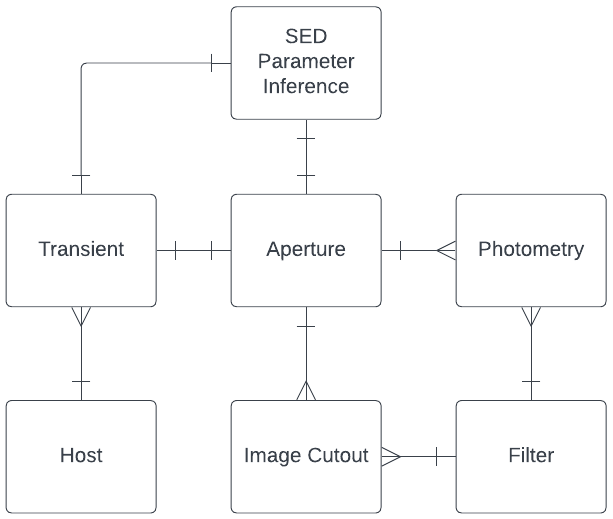}
    \caption{Simplified schema showing how {\tt Blast} connects transients to host galaxies, photometric apertures (local or global), filters, image cutouts, and ultimately SED inference.  Each box represents a database table, with three lines connecting to a box indicating ``many" and a single perpendicular line indicating ``one".  For example, each filter can be associated with many image cutouts, while a given image cutout is associated with only one filter.}
    \label{fig:schema}
\end{figure}

The {\tt Blast} data model is shown in Figure \ref{fig:schema}.  The goal of the data model is to store every astrophysical transient and its associated metadata (via the {\tt Transient} table), and connect it to its associated host-galaxy information ({\tt Host}), the photometry of both the host galaxy as a whole and a circular aperture near the transient location ({\tt Photometry}), and its derived physical parameters ({\tt SED Parameter Inference}).  Simultaneously, it is connected to the data sets needed to provide that information ({\tt Aperture}, {\tt Image Cutout}, and {\tt Filter}).  Finally, the tasks are managed by {\tt Task} and {\tt TaskRegister} tables, which record for each transient the tasks that have been run and the status of those tasks.

\subsection{Processing Workflow}
\label{sec:processing}

\begin{figure}
    \centering
    \includegraphics[width=3.4in]{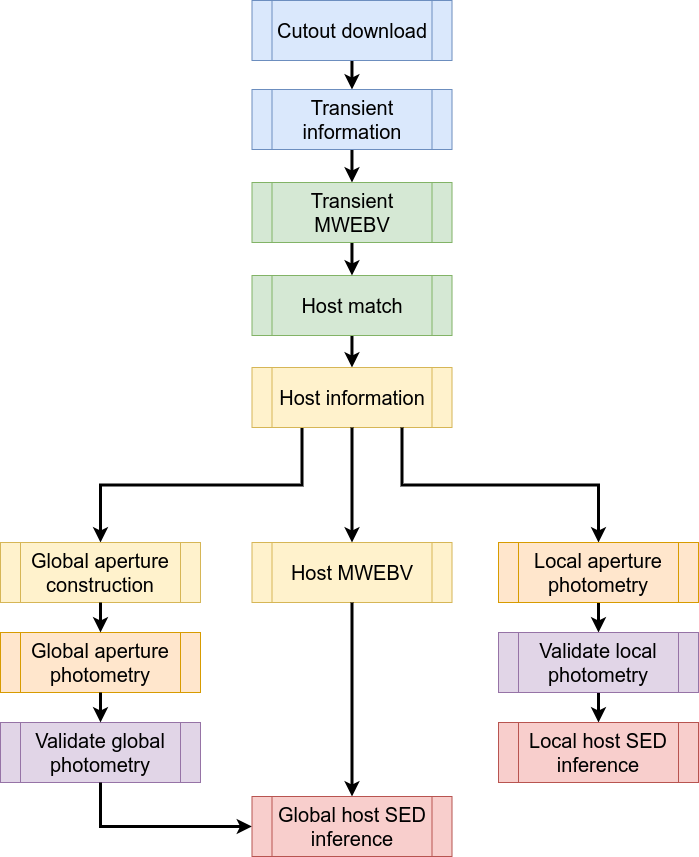}
    \caption{Diagram of the transient workflow directed acyclic graph (DAG).  Tasks are described in Section \ref{sec:processing}.}
    \label{fig:dag}
\end{figure}

Celery Beat periodically executes a task that polls the Transient Name Server,\footnote{\url{http://wis-tns.org/}.} launching processing workflows for each newly detected transient. Each workflow is comprised of tasks that Celery schedules to run asynchronously across a pool of workers, maximizing utilization of the available compute resources and completing workflows as quickly as possible. The workflow is implemented using Celery's native ``canvas'' system, which ensures that tasks in a workflow are executed according to the specified directed acyclic graph (DAG) of task dependencies (Figure \ref{fig:dag}). Augmenting the Celery workflow system is a custom state management system, whereby completed tasks are marked as ``processed'' such that subsequent executions of the same workflow do not waste time reprocessing the same data. This supports efficiently resuming prematurely aborted workflows. For example, a global host-galaxy SED-fitting task will be launched only if the status of that task is ``unprocessed" and host-galaxy photometry has been performed successfully\footnote{Note: The term ``task" is overloaded. Sometimes the word ``task'' refers to a defined data processing \textit{type} (e.g., {\tt Host SED fitting}), and sometimes it refers to an \textit{instance} of such a task's execution. Hopefully context will make the distinction clear.}.

\subsubsection{Transient Data Ingestion}

{\tt Blast} ingests new astrophysical transients by querying the Transient Name Server\footnote{\url{https://www.wis-tns.org/}.} (TNS) API for transients that have recently been discovered or updated.  This query runs every few minutes by default, but can be modified to run at any cadence. Additional tasks can be defined by extending the {\tt TaskRunner} class and appending the new class to the list of periodic tasks; thus a developer could add modules to query data from, for example, ZTF (likely via, e.g., the Alerce, Lasair, or ANTARES brokers; \citealp{SanchezSaez21,Smith19,Matheson21}) or the Rubin Observatory's LSST.

New transients may also be added manually by authenticated {\tt Blast} users in one of two ways: the first is by entering the name of a TNS-reported transient, triggering a query of the TNS API to ingest that transient's metadata and initialize the transient workflow; the second is by specifying the name, right ascension, declination, redshift, and class of a transient as an uploaded document. The second method supports a broader range of non-public transients or galaxy science.

TNS data includes coordinates of the transient and redshifts derived from transient spectra, when available.  We also obtain Milky Way reddening estimates from \citet{Schlafly11} via the {\tt dustmaps} python package \citep{Green2018}.

\subsubsection{Downloading Imaging Data}

The {\tt Blast} architecture allows individual instruments and filters to be defined via a file that is ingested into the database on startup.  The user must define image characteristics like the filter name, survey, units of the image (usually counts, or counts per second), effective wavelength, pixel size, zeropoint (when set to a standard value), and AB offset.
With this information,  {\tt Blast} automatically downloads images, measures photometry, and performs SED fitting for these user-defined instruments\footnote{In practice, adding new filters or instruments will also require re-training the SBI$++$ model for SED parameter estimation; see Section \ref{sec:sed}.}.

In this work we use GALEX \citep{Martin2005}, SDSS \citep{York2000}, Pan-STARRS \citep{Chambers2016,Flewelling20}, DES \citep{Dey2019}, 2MASS \citep{Skrutskie2006}, and WISE \citep{Wright2010} imaging.  Images are downloaded using {\tt astroquery} \citep{Ginsburg2019} via 
MAST for GALEX images, the Pan-STARRS image cutout service for Pan-STARRS data,\footnote{\url{https://ps1images.stsci.edu/cgi-bin/ps1cutouts}.} IRSA for WISE and 2MASS,\footnote{\url{https://irsa.ipac.caltech.edu/frontpage/}.} NOIRLab's Astro Data Lab for DES \citep{Fitzpatrick14,Nikutta20}, and SDSS DR16 via \url{https://data.sdss.org/sas} for SDSS imaging downloads \citep{Ahumada20}.  We require images that have accompanying exposure times or weight maps to estimate reliable photometric errors.

Developers can add additional instruments/filters to the data model.  Because there is no catch-all service for FITS images, developers
need to also add a custom image download function to incorporate new instruments.\footnote{See the developer guide at \url{https://blast.readthedocs.io} for more information.}

\subsubsection{Transient--Host Galaxy Matching}

Transients are matched to their host galaxies using the Galaxies HOsting Supernova Transients \citep[GHOST;][]{Gagliano2021} software. By default, GHOST uses a modified directional light radius method \citep[DLR;][]{2016Gupta_DLR} to identify the most likely host galaxy in the Pan-STARRS 3$\pi$ survey within 60$\arcsec$ of the queried transient (if not found by name or coordinates in the associated GHOST catalog). SkyMapper at Siding Spring Observatory \citep{2024Onken_SkyMapper} is used to identify southern-hemisphere hosts, for its similar filter set and depth to Pan-STARRS (SkyMapper has a 5-$\sigma$ single-epoch detection limit of $m_g\sim$23, comparable to PS1's stacked photometry). Candidate extended sources are pre-filtered using PS1 quality flags and a random-forest classifier optimized for star-galaxy separation. The PS1 source from the stacked source catalog with the lowest DLR value is selected as the host galaxy; DLR computation uses the Kron radius of the source \citep{Kron80}, the elliptical radius computed from the first moment of the image, in the filter with the highest signal-to-noise ratio. If the closest source has DLR$>$1 (indicating that the transient is beyond the radius of \textit{any} source), sources classified in the NASA/IPAC Extragalactic database\footnote{The NASA/IPAC Extragalactic Database is funded by the National Aeronautics and Space Administration and operated by the California Institute of Technology.} (NED) as galaxies are preferentially associated. We have found empirically that this update decreases the rate of misassociations.

To address the problem of source shredding (galaxies that become split into multiple resolved segments) for bright and extended low-redshift galaxies, the GHOST software has also been modified to first conduct a DLR association within the GLADE+ catalog \citep{2022Dalya_GLADE}, which contains 22.5M galaxies and is complete to a luminosity distance of 47~Mpc. If no association can be made to a GLADE+ source, the full pipeline is run among PS1 sources. This triaging approach significantly reduces runtime for local sources, from $\sim$60s per PS1 association to $\sim$10s per GLADE+ association (though we note that a significant fraction of the PS1 runtime is added overhead from remote queries that would be unnecessary if the catalogs existed on disk). GHOST also now takes advantage of hierarchical classifications provided by the SIMBAD astronomical database \citep{2000Wenger_SIMBAD} to ensure that final associations are not made to galaxy sub-structures. 

GHOST additionally provides photometric redshifts via {\tt Easy PhotoZ} for each host galaxy estimated from a neural network (multilayer perceptron) and spectroscopic redshifts from NED, when available.  See Appendix A of \citet{Aleo2022} for additional details on the model.

\subsubsection{Aperture photometry}

We perform aperture photometry by first constructing ``local" and ``global" apertures for each transient in {\tt Blast}. For local apertures, we center a circular aperture with 2~kpc radius at the transient location.  To compute the physical size of that radius we use the photometric or spectroscopic redshift to define the physical size of the aperture assuming a standard $\Lambda$CDM cosmology (H$_0 = 70~{\rm km ~s^{-1}~Mpc^{-1}}$, $\Omega_m = 0.315$; \citealp{Planck18}).  For transients without redshift information, we choose to not provide local host-galaxy properties within some standard aperture as the interpretation of that information would be unclear (if an updated redshift is sent to TNS, all tasks are re-processed).

In a separate validation stage, we ensure that each filter/instrument has sufficient resolving power at the angular size of 2~kpc in a given galaxy.  Conservatively, we require that the PSF FWHM of a given instrument is less than this 2~kpc radius.  For this reason, we caution that the higher the redshift, the fewer the photometric filters that are used to compute local properties.
 
For global apertures, we use the {\tt Astropy Photutils} \citep{Bradley2022} package to find and construct elliptical apertures for each host galaxy identified by GHOST.  {\tt Photutils} constructs an image segmentation map of detected sources ($>3\sigma$ from all connected pixels) and constructs Kron apertures from that map.  We use optical data (preferentially $g$ or $r$ bands from SDSS, Pan-STARRS, or DES, with 2MASS $J$ as a last resort) to compute this global aperture.  We require that an object have at least 10 connected pixels to be considered a distinct object (the ``deblending" theshold), a relatively large value that results in more low-redshift galaxies being treated as a single object instead of being deblended into multiple sources.  Despite this, we still have occasional difficulties with under-estimating the proper aperture size in the case of very bright, nearby galaxies.

\begin{figure*}
    \centering
    \includegraphics[width=7in]{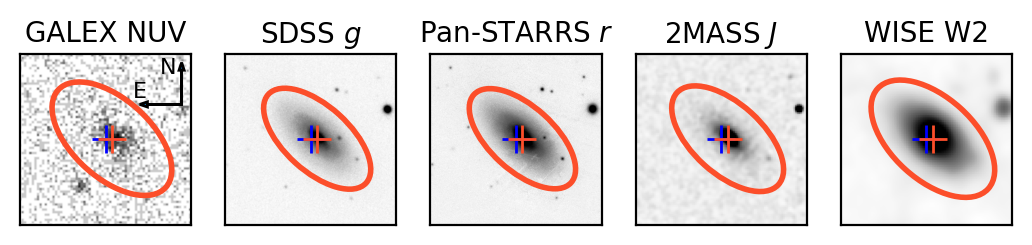}
    \caption{An example of PSF-matched apertures (red ellipses) for SN~2010H for a selection of available cutouts from the UV to the NIR. Each image is 50\arcsec$\times$50\arcsec\ in size. The matched host galaxy and SN~2010H's position are marked as a red and blue cross, respectively.  The directional arrows in the left-most panel have lengths of 20\arcsec.}
    \label{fig:aperture_matching}
\end{figure*}

Because the PSF size of each filter and instrument varies significantly, we make an approximate adjustment to the Kron radius for a self-consistent flux measurement across all images (Figure \ref{fig:aperture_matching}).  We increase or decrease each aperture by the difference in standard deviation of the seeing between the reference image used to compute the aperture (typically, $g$ band) and the image of interest.  While a perfect correction for the effect of the convolution of the galaxy light with each PSF would be morphology dependent, and therefore difficult in practice, this approach gives us an approximate adjustment that treats the outer edge of galaxy light approximately the same across wavelengths.

The uncertainties on each image are assumed to be Poisson-noise dominated and we use the calibrated zeropoints for each image provided by its individual survey.  For WISE filters, uncertainties are correlated pixel-to-pixel, and we therefore apply a noise correction term as a function of the number of pixels in an aperture.\footnote{Computed by the WISE team and described at \url{https://wise2.ipac.caltech.edu/docs/release/prelim/expsup/sec2_3f.html}.}

For each aperture, we check whether neighboring sources are likely to contaminate a given photometric measurement.  We first generate segmentation maps from both an optical image and the image on which photometry is being performed.  If the segmentation map flags multiple objects within the aperture in either image then the photometry is considered to be ``contaminated" and is not used in fitting the SED.  Due to the complexity and intrinsic assumptions required to correct for contamination within an image, we do not attempt to model and correct for contaminating sources in the current work.  The effect of this treatment is that images with larger PSFs (e.g., GALEX, WISE) are more likely to have contaminating sources and are frequently excluded from SED fitting.

In rare cases, very low-redshift galaxies are deblended into multiple sources that overlap with our aperture, and therefore every global photometric detection is considered contaminated.  In these cases, we check to see whether all photometric measurements are flagged as contaminated, and if so we consider this a likely case of a low-redshift galaxy that is not truly contaminated by neighboring objects.  We then issue a warning to the user (visible via the web application results pages --- see Section \ref{sec:app}) and include all photometry in the SED fit.

\subsubsection{Host SED fitting}
\label{sec:sed}


SED fitting is performed using {\tt Prospector} \citep{Leja2017,Johnson2021}.  Within {\tt Prospector},
we use the MIST stellar isochrones \citep{Choi2016,Dotter2016}, and the MILES stellar spectral library \citep{SanchezBlazquez2006} in the Fast Stellar Population Synthesis (FSPS) code \citep{Conroy2009, Conroy2010ascl, Conroy2010}. A non-parametric SFH is characterized by mass formed in seven logarithmically spaced time bins \citep{Leja2017}, with a continuity prior to ensure smooth transitions between bins \citep{Leja2019}. Nebular emission is accounted for using a pre-computed Cloudy grid \citep{Byler17}. Dust is modeled by two components \citep{Charlot00}, with a power-law parameterization to a \citealt{Calzetti00} curve \citep{Noll09}.
Emission from an AGN torus is parameterized as a simplified version of the CLUMPY model \citep{Nenkova08a}, adding the normalization and dust optical depth as two additional free parameters \citep{Leja18}. Dust emission \citep{Draine07} is incorporated into all fits. The full set of parameters are described in Appendix \ref{app:priors} and \citet[their Table 1]{Wang23}.\footnote{They are also included in the {\tt Blast} documentation: \url{https://blast.readthedocs.io/en/latest/usage/sed_params.html}.}  Adopting a full Bayesian approach means that when a parameter is not constrained by the available data, its uncertainty is propagated into the rest of the inferred parameters (i.e., it is essentially treated as a nuisance parameter and is marginalized over).


In tests on our data we found that obtaining posteriors using the Prospector-$\alpha$ model with nested sampling \citep[via {\tt dynesty};][]{Speagle2020} could typically take $>24$ core-hours of computation. 
Replacing the generation of stellar population models \citep{Bruzual2003,Conroy2010} with an emulator \citep{Alsing20,Hahn23,Mathews23,Thorp24} decreases the computation time to $\gtrsim 15$~mins (for a large-scale application, see \citealt{Wang2024:sps}). A number of other techniques have also been proposed, including gradient-based samplers \citep{Duane87,Hoffman11}, and samplers based on a normalizing flow \citep{Karamanis22,Wong23}. Here, 
we use the simulation-based inference (SBI) method presented in \citet{Wang23} as SBI$++$. SBI$++$ is a neural posterior estimator that quickly returns full posteriors on the full set of model parameters.  SBI$++$ is constructed using Masked Auto-regressive Flow \citep{Papamakarios17}, as implemented in the {\tt sbi} Python package \citep{Greenberg19,TejeroCantero20}. As described in \citet{Wang23}, the model has 15 blocks, with each block having two hidden layers and 500 hidden units.  

The SBI$++$ method expands upon SBI by using Monte Carlo (MC) sampling to replace missing bands and out-of-distribution uncertainties.  We train our model on 1.5 million simulated galaxies; 2/3 of galaxies have parameters randomly drawn from priors representing typical values for e.g., stellar mass, star formation history, metallicity, etc., while 1/3 use flat priors spanning a wide range of values.  The full prior distributions are listed in Appendix \ref{app:priors}. 
 Simulated redshift is drawn from a flat distribution spanning $z = 0.0$ to $z = 0.2$ (note that redshift will be fixed to a spectroscopic or photometric redshift in the SED fitting stage as discussed below).  This redshift range encompasses most bona fide SNe being discovered and reported by current surveys.

From these parameters, FSPS/Prospector produces model photometry. We add Gaussian noise to that photometry by using the mean signal-to-noise ratio (S/N) as a function of magnitude measured for each photometric filter in our data set.  We then train a neural density estimator on the simulated training data, which will then be used to infer the host-galaxy parameters of our real data.  Training each {\tt Blast} model took approximately four days on a single Intel ``Sapphire Rapids" CPU.

In fitting the data, we use either a spectroscopic or photometric redshift, where available; if no spectroscopic or photometric redshift exists, we do not attempt to perform SED parameter estimation.  Although future work will include a supplemental trained SBI$++$ for simultaneously estimating photometric redshifts in {\tt Blast} \citep[e.g.,][]{Wang23}, the current version uses the GHOST algorithm for photo-$z$s prior to SED fitting.
While this means that the uncertainties on the redshift are not propagated into the inferred stellar population parameters, the most scientifically interesting transients tend to be spectroscopically classified (i.e., the uncertainties on these redshifts are small), and approximately 30\% of host galaxies ingested into {\tt Blast} to date have reliable spectroscopic redshifts. We therefore employ a fixed-redshift version of SBI$++$ for the initial version of {\tt Blast}.

As is implemented in SBI++, if there are missing bands in the observed photometry, we find 50 nearest neighbors in the simulated training data to the real data. For each MC sample we draw 50 posterior samples (i.e., 2500 posterior samples are used in total).  Although the fluxes of non-detections are measured and recorded, we do not use $<3\sigma$ detections for SED fitting in the current work because the SBI$++$ framework is trained on magnitudes to limit the dynamic range of the values. 

The full fitting process typically takes between $5-10$ core-minutes, and is capped at 2 core-hours for an occasional galaxy that may have poorly estimated photometry or have parameters not well-represented by the simulated training data.  In practice, we find that this occurs just 2\% of the time for global SED estimation, and 0.6\% of the time for local estimation. We find that these failures are frequently the result of data quality issues that affect the photometry.

\begin{figure}
    \centering
    \includegraphics[width=3.4in]{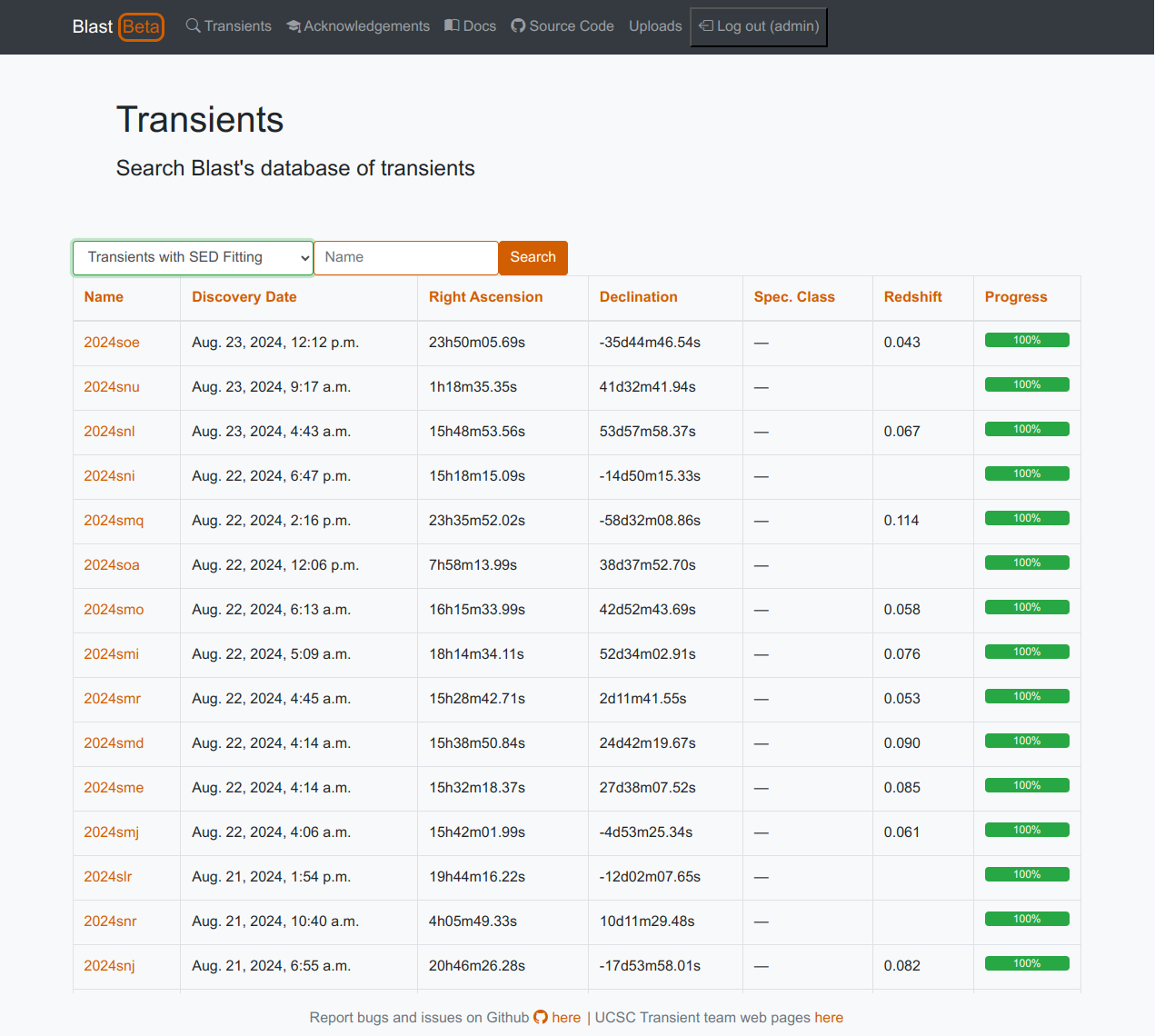}
    \caption{An example of the {\tt Blast} transient processing queue, using a subset of recent and archival transients. For each transient, basic information such as transient name, discovery date, sky position, redshift, transient class, and {\tt Blast} live processing progress are shown. Users can search for transients by name or by some pre-loaded common filters such as transients with their host-galaxy spectral energy distribution (SED) fitting completed, as shown in this example screenshot.}
    \label{fig:transient_list}
\end{figure}

\begin{figure*}
    \centering
    \includegraphics[width=0.75\linewidth]{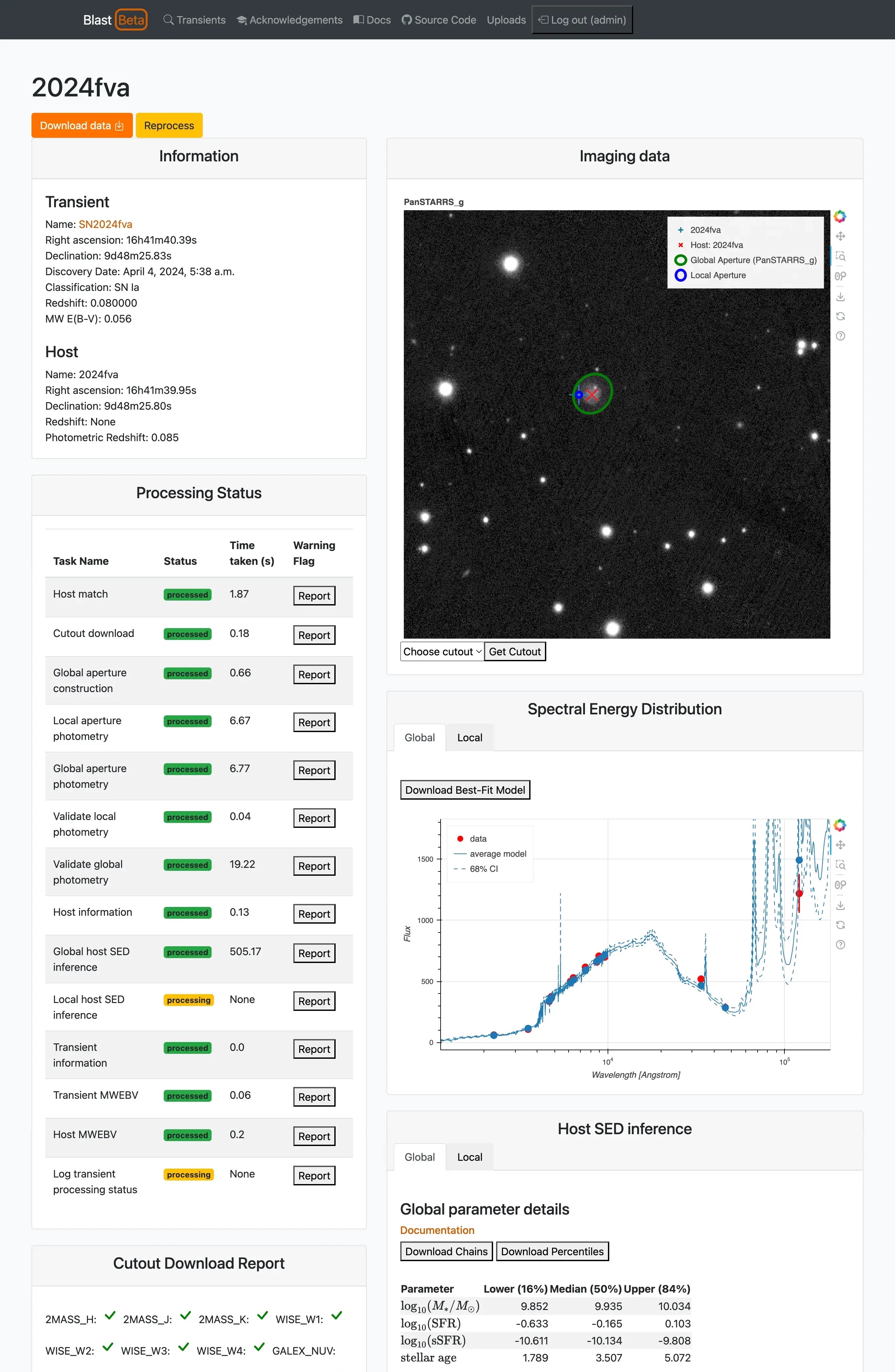}
    \caption{The {\tt Blast} transient results page for SN~2024fva, including basic information, task statuses, cutout images, photometry, best-fit SED models, and SED parameter estimation. The imaging data panel (top right) allows users to select from all available cutout images acquired for the transient. Both the imaging data and spectral energy distribution panel (middle right) are plotted using {\tt Bokeh} \citep{Bokeh18:bokeh} which allows the user to interact with the data by means of panning, zooming, and downloading plots. The user can also download SED best-fit models and posterior parameter estimations from this page (lower right), and flag issues with {\tt blast} processing via the report button in the processing status panel (left side). Finally, users can also download the entire {\tt Blast} dataset for a transient by clicking the download data button in the top left corner. }
    \label{fig:transient_results}
\end{figure*}


\subsection{Web Application}
\label{sec:app}

The {\tt Blast} web application is designed to provide each transient's processing status, images, photometry, and SED parameters as transparently as possible.  The home page provides a basic description of {\tt Blast} and basic summary statistics of transients ingested and completed.  

Figure \ref{fig:transient_list} shows the ``Transients" page, containing a table of all transients ingested into {\tt Blast}.  For each transient, the table gives coordinates, type and redshift (when available), and a status bar indicating progress of the transient processing workflow.  It includes a search function and drop-down menu allowing the user to see transients with host-galaxy matches, transients with photometry, transients with SED fitting results, or transients that have successfully completed all stages.

The transient ``Results" page (Figure \ref{fig:transient_results}) summarizes all available information for a given transient.  This includes the following:

\begin{enumerate}
    \item Basic transient information, including coordinates, type, and redshift.
    \item Basic host galaxy information, including name, coordinates, redshift, and photometric redshift (if available).
    \item The status of all tasks.  Typical status messages are ``unprocessed," ``processed," ``processing," or ``failed," and occasionally more informative messages are used to explain why a stage did not complete; for example, ``redshift too high," or ``no local phot" are typical errors when the redshift is out of range for our trained SED-fitting model or the PSFs of all instruments are too broad to probe a 2~kpc physical scale.
    \item Interactive plotting of each archival host-galaxy image (the default is Pan-STARRS $g$, with alternatives viewable using a drop-down menu).  Figures include both the transient and host-galaxy location as well as the local and global apertures for which photometry is being measured.
    \item Local and global photometric measurements, including magnitude and flux ($\mu$Jy).
    \item Interactive figures with the photometric data. SED model spectra corresponding to the 16th/50th/84th percentiles from posteriors are shown with dashed lines.
    \item The 16th, 50th, and 84th percentile measurements of all Prospector-$\alpha$ parameters as well as the following derived parameters: surviving stellar mass, star formation rate (SFR) and specific star formation rate (sSFR) averaged over the last 100~Myr (the full star-formation history is also available), and mean mass-weighted stellar age.
    \item Options to download the data.  The ``Download data" button provides all metadata available for a given transient, including photometry and SED parameters.  There are individual buttons for downloading the best-fit SED model, the 16th/50th/84th percentile parameter values for every parameter in the Prospector-$\alpha$ model, and the full posterior chains.  A Django Rest Framework API provides more fine-grained information to the user and can be accessed programatically as described in our documentation. 
\end{enumerate}

The uploads page allows users with a {\tt Blast} account to upload transients not currently in the database either by IAU name or by a comma-separated list of object name, right ascension, declination and (optionally) redshift and transient type.  For the former method, information is ingested from TNS and for the latter, no additional information is necessary before {\tt Blast} can begin processing.  We restrict this page to registered users to limit the rate of transients ingested to the database, but for both public and local installations, customizable uploads will make {\tt Blast} more useful for both older transients and non-transient science cases that require galaxy characterization.

Finally, the API provides access to all data associated with the data model.  The Rest Framework API can be accessed via the web interface as well as programmatically.  Simple filters help the user sort through tables as specified in the documentation; for the {\tt Transient} table this includes filters on transient name and redshift, and for most other tables, filters allow the user to easily select an associated transient or filter by filtering on foreign-key relationships as described in the documentation.

\subsection{Deployment}
\label{sec:deployment}

There are two methods for deploying {\tt Blast}: Docker Compose and Kubernetes. The Docker Compose method is ideal for local application development, although it would also suffice as a production instance on a single machine. In this mode of deployment, the system components described in Section \ref{sec:architecture} are deployed as local containers with network-based interactions that closely resemble the way the components are deployed and interact with one another in the Kubernetes environment. By aligning these two deployment methods as much as possible, there is minimal overhead involved in translating newly developed features and bug fixes from the development to the production environments. 
The production deployment is driven by a Helm chart crafted to exploit the scalability and high availability of a Kubernetes cluster.

The public deployment of {\tt Blast} can be found at \url{http://blast.scimma.org/} with full documentation at \url{https://blast.readthedocs.io/}.  {\tt Blast} is deployed on resources provided by the NSF ACCESS program \citep{NSFACCESS} as part of the Scalable Cyberinfrastructure to support Multi-Messenger Astrophysics (SCiMMA) project to aid in the characterization of the host galaxies of multi-messenger sources such as kilonovae.  We welcome new developers and contributors; please read the developer guide in our documentation to get started.\footnote{\url{https://blast.readthedocs.io/en/latest/developer_guide/dev_getting_started.html}.}

\section{Science verification}\label{sec:verification}

In this section, we present the results of several tests to ensure that {\tt Blast}'s data products are reliable and consistent with previous results in the literature.  These include host-galaxy matching, Kron aperture estimation, flux/magnitude measurements in each filter, and SED parameters.  

We note that {\tt Blast} has occasional failures in each of these categories, and we attempt to describe potential failure modes below.  Although these occasional issues might not be obvious if the user downloads data in bulk, they should all be apparent when viewing the individual web pages in the {\tt Blast} application.  Additionally, if a user has indicated that a given stage has failed to complete satisfactorily, this information can be obtained via the API.  Users also have the option of ``reporting" the result of a given stage as erroneous and, in the case of singular errors like a third-party service going offline, registered users can request to re-process the entire transient.

\subsection{Host Galaxy matching and Kron aperture determinations}

One of the first processing tasks that is run on transients ingested into {\tt Blast} is finding a host-galaxy match. To perform this matching we use the GHOST software via the astro\_ghost python package\footnote{\url{https://github.com/uiucsn/astro_ghost}.}.

\begin{figure}
    \centering
    \includegraphics[width=0.9\columnwidth]{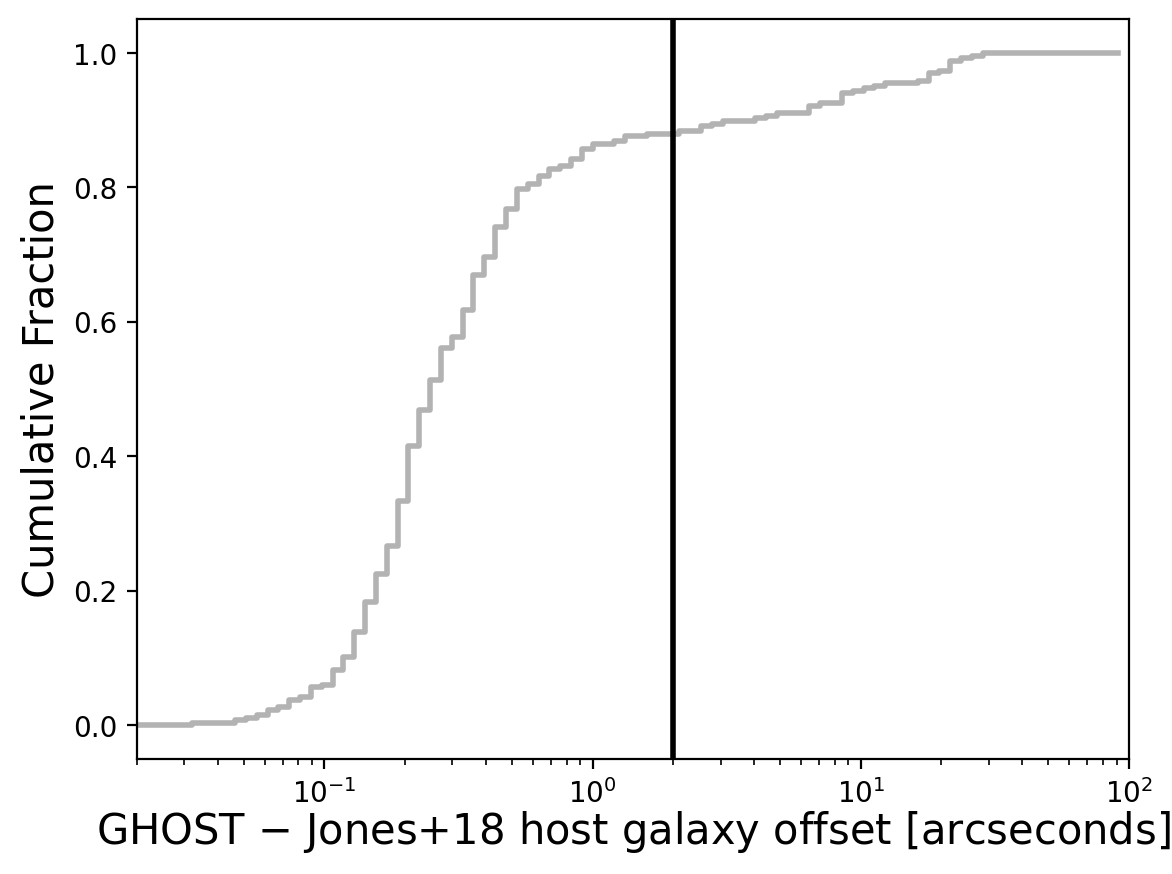}
    \caption{Cumulative distribution of offsets between GHOST and the \cite{Jones2018} host galaxies from the sample of 273 SNe\,Ia that were checked by eye in \cite{Jones2018} (two ambiguous matches were removed).  We note that \citet{Jones2018} ran SExtractor on an individual Pan-STARRS image to find host centroids for the best-fit ellipse (with by-eye verification), while GHOST uses the mean position from all Pan-STARRS images, resulting in typical centroid differences of $\sim$1-2~pixels that are unrelated to the quality of the choice in host. The vertical line indicates a $2$-arcsecond offset. Approximately $88\%$ of SNe have consistent host-galaxy coordinates within 2 arcsec (denoted by the vertical line), and for an additional $\sim$5\% GHOST has identified the same host galaxy as \citet{Jones2018} but has offset coordinates for the host center due an extended and/or irregular host galaxy.  GHOST identifies a different host galaxy 6.6\% of the time (Figure \ref{fig:GHOST_match_mosaic}).}
    \label{fig:GHOST_jones_sep_dist}
\end{figure}

\begin{figure*}
    \centering
    \includegraphics[width=\textwidth]{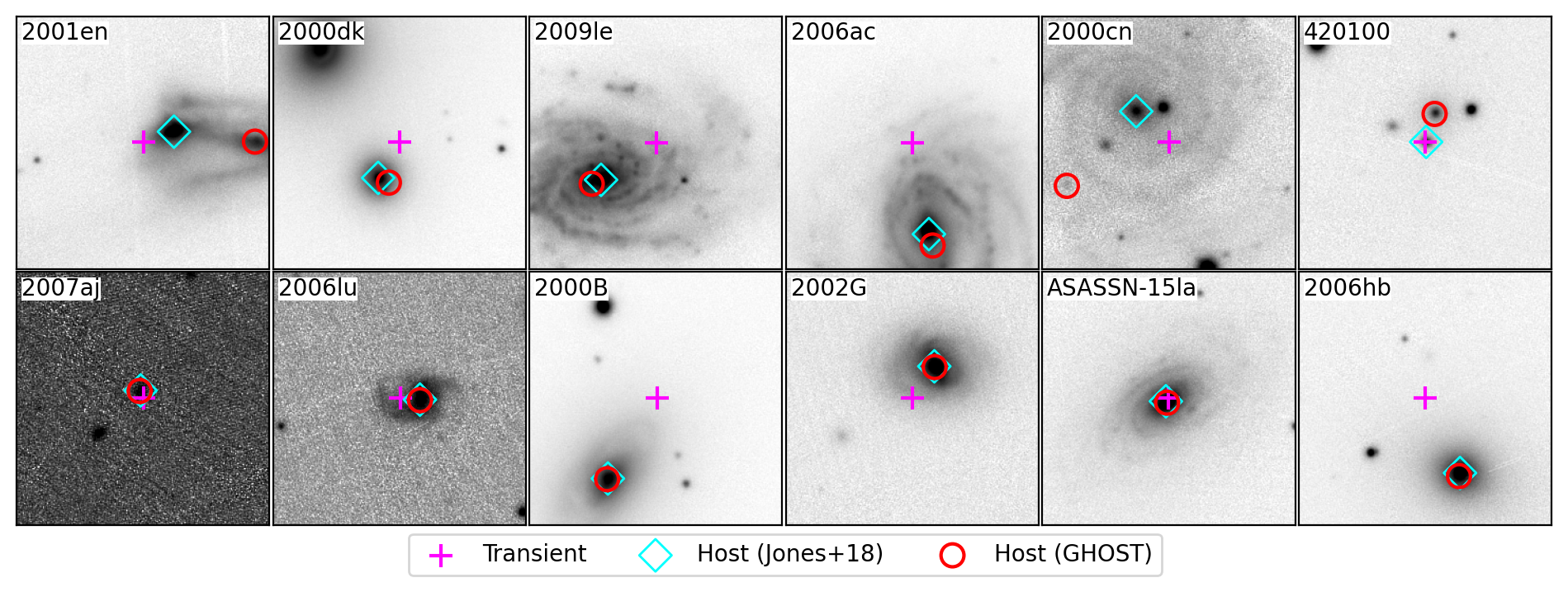}
    \caption{Examples of GHOST \citep{Gagliano2021} host-galaxy associations compared with \cite{Jones2018} (Jones+18). The top row shows associations where GHOST and \cite{Jones2018} disagree (host galaxy $>2.0$ arcsecond offset), and bottom row shows agreement (host galaxy $<2.0$ arcsecond offset). All images are Pan-STARRS 65$\times$65 arcsecond g-band cutouts.  Most associations shown on the top row are differences in centroids that will result in the same final host-galaxy apertures, but SN~2000cn shows a case where GHOST is definitely incorrect and SN~420100 shows a case where GHOST is likely incorrect.}
    \label{fig:GHOST_match_mosaic}
\end{figure*}

We validated GHOST's performance against the host-galaxy matches of the low redshift ($z<0.1$) sample of $273$ SNe\,Ia from \cite{Jones2018} (Figures \ref{fig:GHOST_jones_sep_dist} and \ref{fig:GHOST_match_mosaic}).  GHOST identifies the same host galaxy in 93.4\% of cases.  Occasional failures result in cases where there is an extended, resolved source galaxy with individual bright components (SN~2000cn), or cases where there are several nearby bright sources (the Pan-STARRS SN 420100).  We note that it is possible that mismatches are more common in non-Ia hosts, especially for transients typically hosted by fainter/smaller galaxies, and we are working closely with the GHOST team to continue reducing the number of host-galaxy mismatches where possible.

\subsection{Kron Aperture Estimation}

\begin{figure*}
    \centering
    \includegraphics[width=7in]{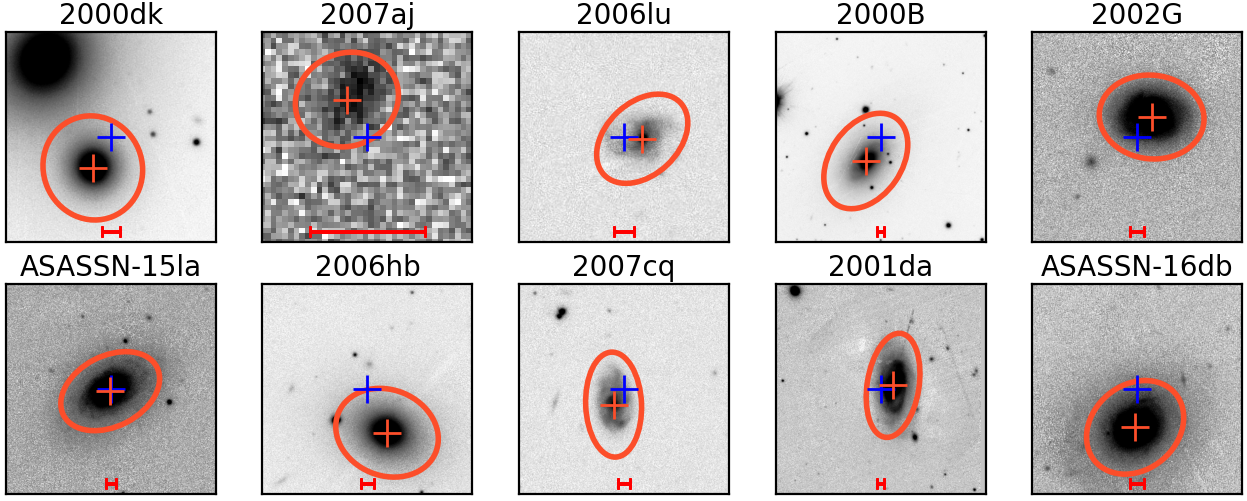}
    \caption{Examples of computed elliptical apertures for photometry from {\tt Blast} using $g$-band images from Pan-STARRS.  The SN position (blue) and the host-galaxy position (red) are shown along with the host-galaxy aperture used for photometry (red ellipses).  Bars show the size of 10\arcsec\ in each image.}
    \label{fig:apertures}
\end{figure*}

Figure \ref{fig:aperture_matching} shows an example of a typical aperture size across bandpasses (using SN~2010H).  Figure \ref{fig:apertures} shows the result of our Kron aperture computation for a subset of Pan-STARRS images.

Our Kron aperture estimation uses the {\tt astropy photutils detect\_sources} algorithm with a 5$\sigma$ threshold and a minimum of 10 connected pixels to find galaxies in an optical image (or 2MASS, in rare cases where no optical images are available).  The {\tt detect\_sources} algorithm constructs a segmentation map to determine the extent of the host galaxy, which we find to be robust in nearly all cases.  We do not perform an additional step of deblending nearby sources, as galaxies are typically well-separated and deblending would result in large nearby star-forming galaxies being separated into multiple objects.  The only failure case we find is that in rare cases involving large, nearby galaxies, the aperture is not always large enough to encompass faint, extended regions.  One compounding factor may be that Pan-STARRS images over-subtract the background near very bright extended sources.

\subsection{Global Galaxy Photometry}

Comparisons of galaxy photometry to catalog photometry are shown in Figures \ref{fig:sv_phot_uvopt} and \ref{fig:sv_phot_ir} for UV/optical and IR photometry, respectively.  For each filter, survey, and telescope, we use publicly provided source catalogs (with the exception of WISE, as described below) and compare to a randomly selected subset of the host-galaxy photometry available from {\tt Blast} at the time of writing.

 The sigma-clipped means are in good agreement with the survey catalogs for GALEX, SDSS, and 2MASS, although 2MASS catalog magnitudes become systematically fainter for sources where 2MASS is unable to detect the wings of the galaxy (compared to the larger, optically-derived {\tt Blast} aperture).  Pan-STARRS is consistent in the $i$-band and slightly offset in the $y$ band, with this offset likely due to the same reason as faint 2MASS magnitudes being systematically fainter: {\tt Blast} apertures are derived from optical photometry and therefore better capture flux in the wings of extended sources. However, at faint magnitudes, {\tt Blast} finds slightly fainter magnitudes than Pan-STARRS on average.  Although we are not sure of the reason for the discrepancy, visual inspection of several outliers appears to indicate that the background in redder bands could be under-subtracted in the Pan-STARRS imaging causing artifically brighter catalog sources.

For WISE, the catalog magnitudes use profile-fitting photometry, which is inaccurate for extended sources; the {\tt Blast} photometry is $\sim$0.5~mag brighter than the catalog photometry.  When we instead compare to WISE forced photometry at the location of SDSS galaxies from \citet{Lang16}, we find that {\tt Blast} photometry is 0.25~mag fainter on average than sources from the unWISE catalog, which could indicate a systematic difference in aperture between the two catalogs and may also impact the number of contaminating sources.  For our baseline comparison, however, we use the method of \citet{Jarrett19,Jarrett23}, which was performed by T.\ Jarrett (private comm.) on the \citet{Jones2018} sample (Figure \ref{fig:sv_phot_ir}).  We see slightly fainter magnitudes in {\tt Blast} (0.15~mag in $W1$), which may indicate that the WISE PSF size is slightly underestimated, but no trend as a function of magnitude.


Upon visual inspection of the 3-$\sigma$ outliers, which are $\sim$5-10\% of the measurements, we find that nearly all are due to complex host-galaxy environments with significant galaxy contamination.  To eliminate these data, we generated segmentation maps to find contaminating objects within a given aperture and remove this contaminated photometry from the set used for SED parameter estimation.
It is also possible that some significant IR emission is beyond the global aperture radius inferred from the optical data and causes systematic differences.


\subsection{SED Parameters}

We compare our transient host-galaxy results to several literature samples to ensure consistent and reliable SED fitting results.  Although relatively few literature samples exist with a low-redshift Prospector-$\alpha$ model specifically, we searched for consistency in parameter measurements using the sources below.  These predominantly use the SN\,Ia sample from \citet{Jones2018}, who analyzed 273 SNe from datasets spanning 1990-2015, but also include a recent Prospector analysis from \citet{Ramsden22}.  SED validation figures are presented in Appendix \ref{sec:app_sed}, as described below:

\begin{figure*}
    \centering
    \includegraphics[width=\linewidth]{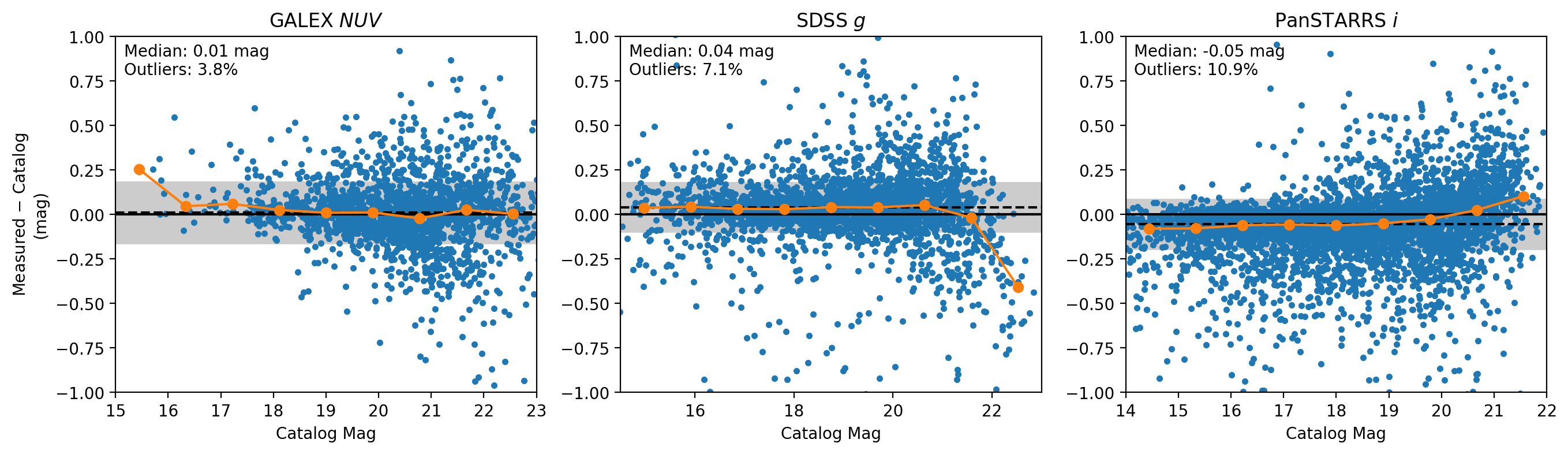}
    \caption{Representative comparisons of {\tt Blast} photometry compared to catalog magnitudes for GALEX, SDSS, and Pan-STARRS (left to right).  The $\sigma$-clipped mean and standard deviation are shown with the dashed black line and grey shaded region, respectively and the binned median is in orange.  The median offset and percentage of 3$\sigma$ outliers are labeled.}
    \label{fig:sv_phot_uvopt}
\end{figure*}

\begin{enumerate}
    \item Global and local masses from \citet{Jones2018}.  This sample used {\tt Z-PEG} \citep{LeBorgne2002} to measure global and local, 2~kpc masses for archival SN\,Ia data.  
    
    \item Global specific star formation rates (sSFRs) from \citet{Jones2018} using LePHARE 
    \citep{Arnouts11}.  
    We clip very low sSFRs as galaxies with negligible star formation unsurprisingly show exceptionally high sSFR scatter.
    \item WISE-based masses and sSFRs for the \citet{Jones2018} sample from T.\ Jarrett (private communication).  
    \item Masses and sSFRs from the \citep{Ramsden22} tidal disruption event sample.  

\end{enumerate}

\begin{figure*}
    \centering
    \includegraphics[width=\linewidth]{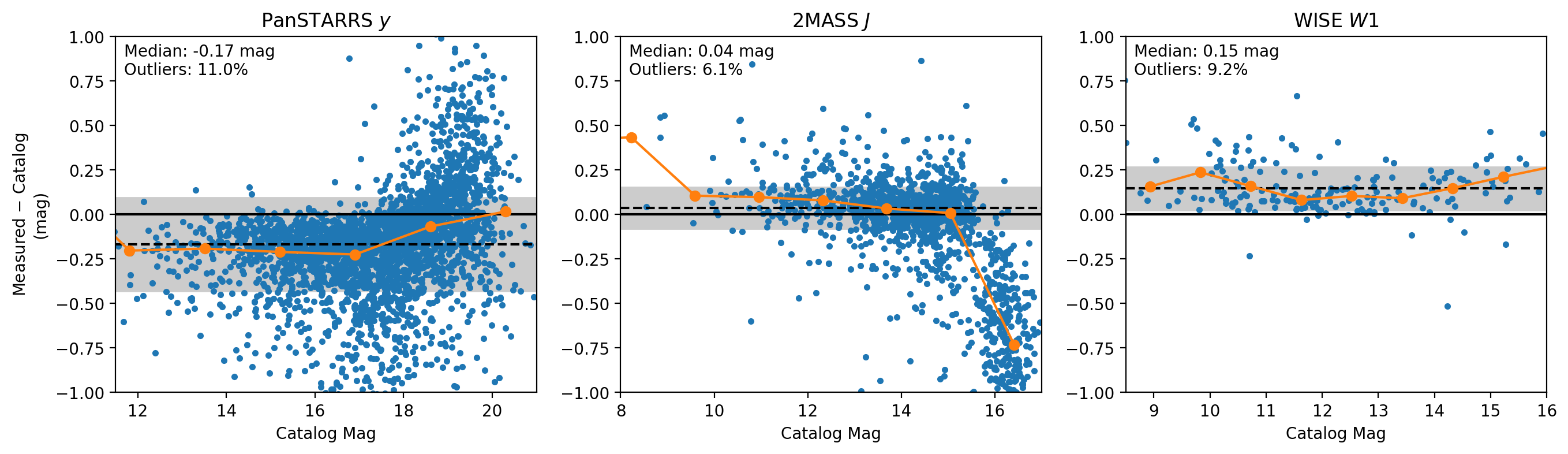}
    \caption{Same as Figure \ref{fig:sv_phot_uvopt}.  Statistics for 2MASS $J$ include only objects with $J < 15$.  WISE $W1$ catalog magnitudes are using the method of \citet{Jarrett23}; there are significant discrepancies (0.5~mag or greater) with the default WISE catalog magnitudes because they assume unresolved sources.}
    \label{fig:sv_phot_ir}
\end{figure*}

Below, we briefly describe these comparisons between {\tt Blast} masses, star formation rates, stellar metallicities, and dust parameters between {\tt Blast} and other sources in the literature.

\subsection{Stellar Masses}

In global mass, there is good agreement between \citet{Jones2018} and {\tt Blast}, with a median offset of 0.07~dex and a dispersion of 0.20~dex.  For local mass, because \citet{Jones2018} uses a 1.5~kpc aperture and {\tt Blast} uses 2~kpc, we apply a correction to the \citet{Jones2018} values that assumes the mass profile is unchanged between the two apertures; this is likely a poor approximation for the lowest-mass regions, which may be more concentrated, and can be seen for the departure from the $y=x$ line at local ${\rm log}(M_{\ast}/M_{\odot}) < 8.5$.  The median offset in local mass is 0.37~dex with a scatter of 0.25~dex.

We also compare to masses derived only from WISE infrared bands, following \citet{Jarrett23}.  There is no obvious trend between {\tt Blast} and WISE-derived masses, with a median offset of 0.060~dex and a dispersion of just 0.12~dex.  We note that \citet{Leja19} found that using a non-parametric star-formation history leads to stellar masses that are larger by $\sim$0.1--0.3~dex as compared to simpler models, which may explain the offset. 
These larger masses are caused by systematically older SFHs inferred from the more flexible non-parametric model, and \citet{Leja19} argue that the improved flexibility in modeling the SFH resolves a long-standing tension between the cosmic SFR density and the stellar mass growth.

The relatively small set of masses from \citet{Ramsden22} have an offset of $-0.043$~dex and a scatter of 0.19~dex compared to {\tt Blast}\footnote{We first convert the {\tt Blast}-reported surviving stellar mass to the total formed stellar mass for this comparison.}; while this difference is somewhat larger than may be expected given that \citet{Ramsden22} use the same Prospector-$\alpha$ model as {\tt Blast}, the photometry, aperture determination, and criteria for choosing which photometric bands are included are different.  There are also likely somewhat different choices in parameters and priors, as well as different star-formation history bins.

We note that additional mass comparisons between {\tt Blast} and the scaling relation from \citet{Taylor11} are published in \citet[their Appendix A]{Peterson24}.  These show largely consistent masses, with no obvious offset or bias across the full dynamic range.

\subsection{Star Formation Rates}

We compare the {\tt Blast} sSFR to the \citet{Jones2018}-derived sSFR from LePHARE as well as sSFRs derived from the WISE mid-infrared bands following \citet{Salim16,Leroy19}.

{\tt Blast} measurements are consistent with measurements from \citet{Jones2018} and WISE-derived measurements, with comparison figures shown in Appendix \ref{sec:app_sed}.  Compared to \citet{Jones2018} the median offset in sSFR is 0.15~dex, although the scatter is a relatively high 0.73~dex.  As can be seen in Figure \ref{fig:ssfrcomp_j18}, {\tt Blast} finds that SNe with the highest sSFR in \citet{Jones2018} appear to be overestimated.  As \citet{Jones2018} is based only upon $ugrizy$ photometry, {\tt Blast} is likely more robust.  We note that the \citet{Jones2018} sSFRs were updated with an improved extinction treatment in \citet[also using LePHARE]{Kim24}, and for these we find qualitatively the same results.  The mean offset is somewhat increased to 0.37~dex and the slightly higher scatter of 0.87~dex is primarily driven by the low-sSFR end.

For WISE, we find slightly lower sSFR from {\tt Blast}, with a median offset of 0.11~dex and an improved scatter, compared to \citealp{Jones2018}, of 0.52~dex.

\subsection{Metallicity and Dust Parameters}

Most other Prospector-$\alpha$ parameters\footnote{A complete list is given in the {\tt Blast} documentation at \url{https://blast.readthedocs.io/en/latest/usage/sed_params.html}.} have larger uncertainties than stellar mass and sSFR.  Although there are relatively few literature studies using Prospector-$\alpha$ at the low redshifts where {\tt Blast} is trained, we show comparisons with \citet{Ramsden22} in Appendix \ref{sec:app_sed} as a general consistency check.  Because we use a full Bayesian approach, we note that in the case where these parameters are unconstrained by the data, {\tt Blast} returns the priors as given in Appendix \ref{app:priors} and these priors are properly propagated into the uncertainties of the other inferred parameters.

In their Table B1, \citet{Ramsden22} report the stellar metallicity ${\rm log}(Z_{\ast}/Z_{\odot})$, and the dust parameters $\tau_{\rm dust,2}$, $n$, and $\tau_{\rm dust,1}/\tau_{\rm dust,2}$.  The $n$ parameter is the power-law index of a modified \citet{Calzetti00} dust attenuation law \citep{Noll09}, the $\tau_{\rm dust,2}$ parameter is the optical depth of diffuse dust, and $\tau_{\rm dust,1}/\tau_{\rm dust,2}$ is the optical depth of stellar birth-cloud dust as a fraction of the diffuse-dust optical depth (this split is based on the \citealp{Charlot00} model).  The ${\rm log}(Z_{\ast}/Z_{\odot})$ and $\tau_{\rm dust,2}$ comparisons are shown in Appendix \ref{sec:app_sed}.

The value of $\tau_{\rm dust,1}/\tau_{\rm dust,2}$ is likely not well constrained and instead returns the prior, which has $\sigma=0.3$; the scatter in measurements is just $\sim$0.06, but the uncertainty on most measurements is $\sim$0.3.  
The $\tau_{\rm dust,2}$ parameter has an offset of $-0.21$ and a scatter of 0.27; the correlation coefficient between this work and the \citet{Ramsden22} values is 0.30.  The other two parameters (shown in Appendix \ref{sec:app_sed}) agree relatively well, with outliers having slight differences in photometry or included bands, but with no obvious indications of poor photometry. There is general consistency in the $n$ parameter, which has an average offset of $\sim$0.1 and a scatter of 0.52.  The large outlier in $n$ is 2019azh, which may be due to the fact that {\tt Blast} does not include the 2MASS bands for this object (and both {\tt Blast} and \citealp{Ramsden22} show poor fits to the WISE W3 band).  For the stellar metallicity, there appears to be a possible systematic disagreement at the high-metallicity end at the $\sim 1.5 \sigma$ level for individual measurements.
However, it is worth pointing out that stellar metallicity is not expected to be tightly constrained by broadband photometry alone, so we do not further explore the origin of such systematic offset.





\section{Science Applications and Current Statistics}
\label{sec:discussion}

\begin{figure*}
    \centering
    \includegraphics[width=0.8\linewidth]{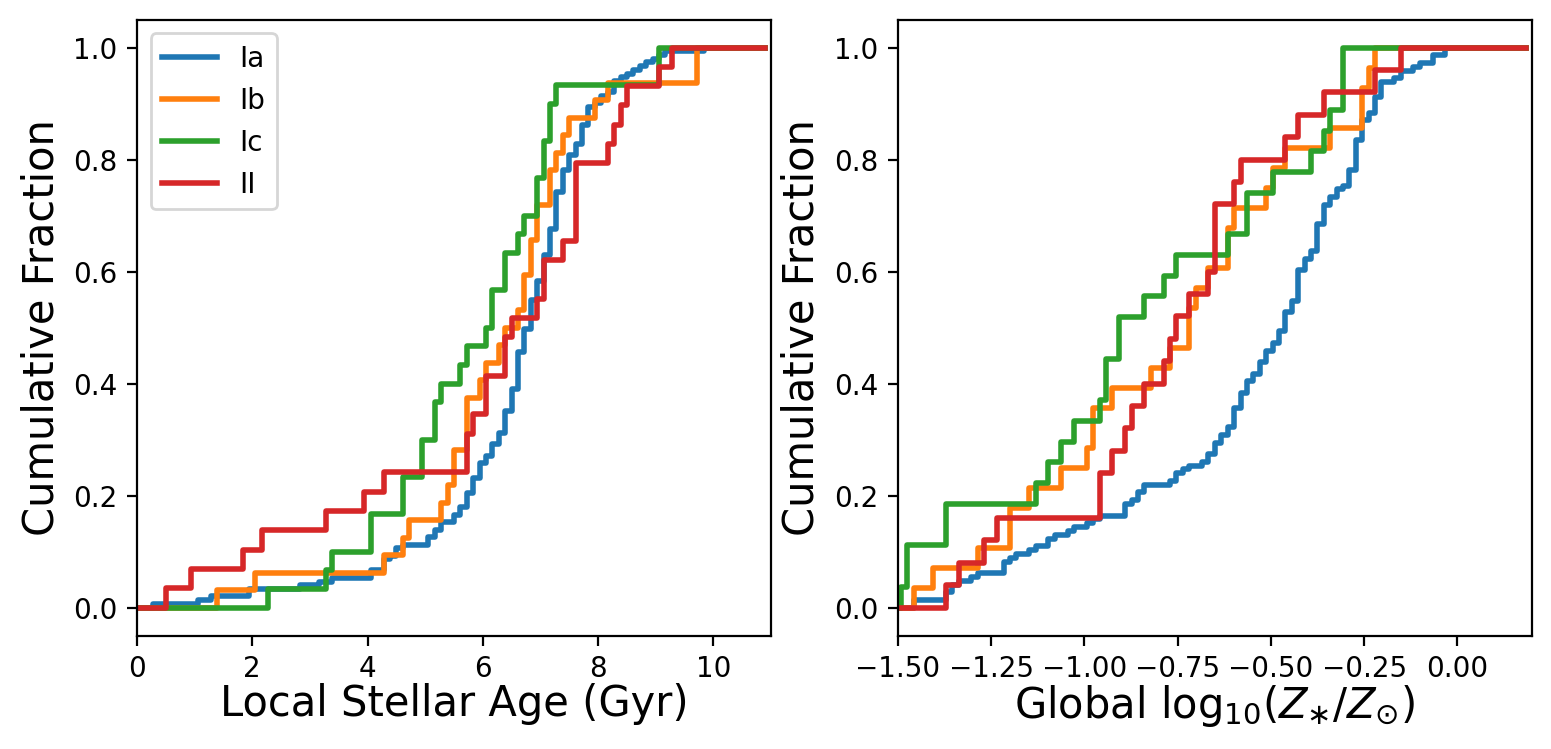}
    \caption{Representative cumulative distribution functions of {\tt Blast}-derived SED properties for SNe\,Ia, Ic, and II.  {\tt Blast} shows that SNe\,Ic prefer locally young regions and globally metal-poor galaxies compared to SNe\,II.  SNe\,Ia are found in the oldest regions and most metal-rich environments on average.  SNe\,Ib and SNe\,II have similar host-galaxy environments, implying similar progenitor properties.}
    \label{fig:sed-summary}
\end{figure*}

\begin{figure}
    \centering
    \includegraphics[width=\linewidth]{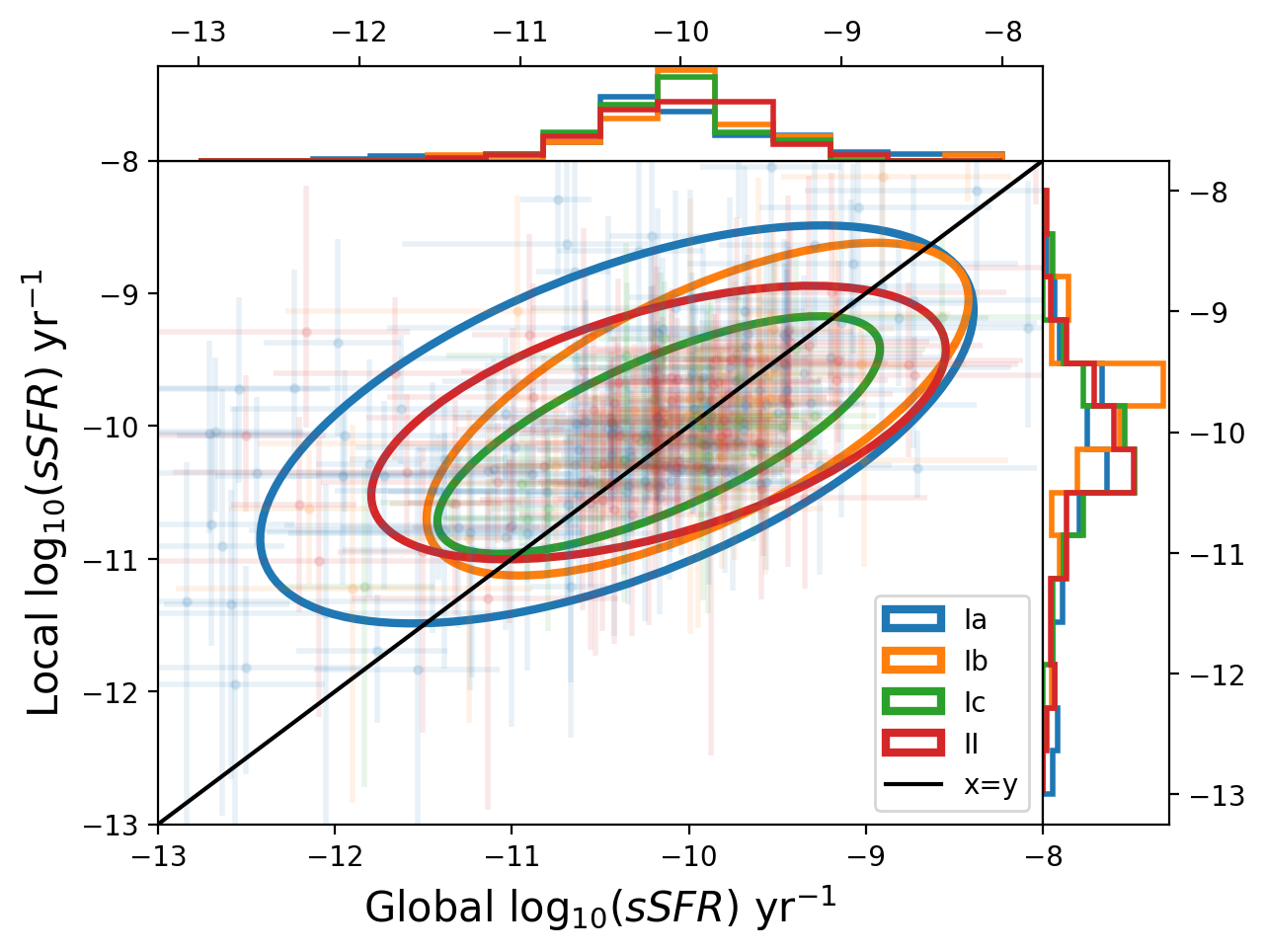}
    \caption{Local versus global sSFR for different subtypes in {\tt Blast}, including SNe\,Ia (blue), Ib (orange), Ic (green), and II (red), with 95\% confidence ellipses.  While most CC\,SNe tend to have similar local and global properties, SNe\,Ia have a greater diversity.}
    \label{fig:localvglobal}
\end{figure}

Currently, {\tt Blast} has ingested more than 13000 transients.  For over 11,000 of these, GHOST has been able to identify likely host galaxies, with most of the remaining transients likely to be stellar flares or hostless SNe (limits on the environments of hostless SNe, e.g., \citealp{Pessi24,Qin24b}, would be extremely interesting but are beyond the scope of the current work).  For $\sim$10,500, host-galaxy photometry was performed; most transients failing at this stage are either lacking redshifts, without which a 2kpc local aperture cannot be computed, and/or are lacking optical data in the Southern sky, which is often necessary to compute an aperture when the host galaxy is undetected in 2MASS.

SED parameters have been successfully estimated for approximately 85\% of recent transients, though we note that past software bugs have prevented SED estimation for some of our sample, with the result that only $\sim$50\% of the full {\tt Blast} database currently has estimated SEDs.  

Currently, the most common failure modes in SED parameter estimation are:

\begin{enumerate}
    \item Redshifts (spectroscopic or photometric) with $z > 0.2$, which is beyond the limit of our training sample.  Sometimes transients are too distant for our current {\tt Blast} training, but more often the photometric redshifts are inaccurate.  Future photo-$z$ estimation within SBI$++$ itself would improve this issue and give more robust posterior estimation.  
    
    Alternatively, a possible future avenue for extending or supplementing the redshift range and parameter space would be the use of emulated SPS 
\citep[e.g.,][]{Alsing20} in combination with conventional posterior sampling. This approach has been demonstrated on large galaxy catalogs \citep{Thorp24}, and can be combined with a calibrated population prior \citep[e.g.,][]{Alsing24} for improved parameter constraints.
    \item The redshift is too distant to resolve a 2~kpc radius for local photometry.  As transients become farther away, we lose the ability to use WISE, GALEX, and 2MASS for SED parameter estimation.
    \item Relatively few photometric bands are available.  Complex host-galaxy backgrounds can cause unreliable aperture photometry, which in turn may cause the photometry to fail our validation cuts.
    \item Extremely large or complex host backgrounds (Figure \ref{fig:blast_extreme_example}).  Automated host identification and aperture estimation for large host galaxies with multiple resolved star clusters, mergers, or other complex morphologies is an ongoing challenge at the lowest redshifts.
\end{enumerate}

\subsection{Science Applications}

The {\tt Blast} application enables host-property inference for individual objects of interest or large-statistics studies of SN types or subtypes.  We envision that key science applications for {\tt Blast} will include:

\begin{itemize}
    \item {\bf The Progenitors of Astrophysical Transients}.  Comparing the environments in which different SN subtypes have been found has shown that stripped-envelope SNe likely originate from binary systems \citep{Smith11,Yoon15}, that SN\,Ic progenitors are more massive than SNe\,Ib or II \citep{Anderson12}, and provided strong evidence that the progenitors of interacting SN IIn maybe be more similar in mass to those of other SN\,II than previously thought \citep{Anderson15}. Host galaxy studies have been also useful in deducing the nature of rapidly evolving transients \citep{Wiseman20b}, Type Ibn supernovae \citep{Hosseinzadeh2019}, SNIc associated with gamma-ray bursts \citep{Modjaz20}, superluminous supernovae \citep{Lunnan15}, and fast blue optical transients \citep{Drout14,Ho20}.  {\tt Blast} unlocks unprecedented statistical leverage for answering many of these questions and provides simple, transparent parameter estimation for individual transients of interest.
    \item {\bf Using Host-Galaxy Properties to Classify Astrophysical Transients}.  Given the thousands of transients discovered each month, classifying and selecting those transients for which additional followup would be scientifically valuable can be critical. Host galaxy based classifiers can provide independent contextual information compared to SN light curve-based classifications given the preferentially different host-galaxy environments of different SN types \citep{Foley2013,Gagliano2021,Qin22,Kisley2023,Lokken2023}.  With real-time SED fitting, {\tt Blast} makes detailed understanding of a transient's host-galaxy properties available quickly.  These SED properties can be used to inform transient follow-up decisions and to train classification algorithms.
    \item {\bf Understanding Cosmological Parameter Measurements from Type Ia SNe}.  Distance measurements from SNe\,Ia can be used to measure the Hubble constant (H$_0$; \citealp[e.g.,][]{Freedman19,Riess22}), the dark energy equation of state \citep[e.g.,][]{Brout22,Vincenzi24}, and the growth of structure in the nearby Universe \citep[e.g.,][]{Boruah20,Stahl21}.  However, interpreting these measurements is complicated by the fact that inferred distance measurements from SNe\,Ia have been shown to correlate with a number of the physical properties of their host galaxies \citep{Kelly2010,Lampeitl2010,Sullivan2010,Rigault2013,Jones2018,Wiseman2020,Brout21,Wiseman23,Dixon24,Grayling24,Popovic24,Ye24}.  In the absence of a consensus for which physical mechanisms are responsible for SN\,Ia--host-galaxy correlations, {\tt Blast} provides a robust, transparent approach for measuring these host-galaxy properties and thereby controlling for the effects of host-galaxy properties on cosmological inference.
\end{itemize}

In Figure \ref{fig:sed-summary}, we show two representative cumulative distributions of SED parameters for different spectroscopically classified SNe currently processed by {\tt Blast}: the mass-weighted stellar age at the location of each SN, and the stellar metallicity for the full host galaxy.  Median uncertainties are $\sim$2~Gyr for the local stellar age and $\sim$0.28~dex for log($Z_{\ast}/Z_{\odot}$).  Using a K-S test we find at high significance that SN\,Ic are have younger stellar ages near their explosion sites than SNe\,Ia and that all core-collapse types have lower stellar metallicities than SNe\,Ia.  While K--S tests for other comparisons indicate that the difference between CC\,SN subtype and SNe\,Ia distributions are not statistically significant, the difference between the medians are 2.4--6$\sigma$ significant for global stellar metallicity and 1.4--3.8$\sigma$ significant for all local age comparisons except for SNe\,Ib and SNe\,II, which are statistically identical.
 With the caveat that observed differences in the median local stellar age of CC\,SN types have marginal (1-2$\sigma$) significance, the distributions shown in Figure \ref{fig:sed-summary} broadly reproduce the expected results: the median SN Ic has a younger stellar age and lower host metallicity than SNe\,Ia or other CC\,SNe.  The SNe\,Ib and SN\,II have similar distributions, which could imply (as pointed out by, e.g., \citealp{Anderson12,Anderson15}) that SN\,Ib are stripped due to binary interactions and have similar progenitor systems to SN\,II.  SNe\,Ia have the oldest stellar ages and the highest mean metallicities.

This figure is just an illustration of {\tt Blast}'s capabilities, and does not account for selection effects, nor does it attempt to use the full sample of available SED properties or the full set of recent SNe from TNS.  Nevertheless, it shows that even some of the more difficult SED parameters to infer robustly can be used to meaningfully probe the nature of transient progenitor systems.

In Figure \ref{fig:localvglobal} we also show local versus global sSFR for different SN subtypes in {\tt Blast}.  While SN\,Ib and Ic largely have similar local and global sSFR, SN\,Ia have a large amount of diversity.  We find that a significant number of SNe\,Ia have higher local than global sSFR; though noise may play a role, we also see from visual inspection that several are in bulge-dominated, largely passive galaxies with residual star-formation activity in the disk (SN\,2000B is one example in our dataset).

As another illustration, we also searched for galaxies near the extremes of the mass and SFR distributions.  We note that the current set of host galaxies in the {\tt Blast} database includes host galaxies of transients reported to TNS beginning in early 2024 as well as transients of interest for certain science cases from our team's collaborators, which may bias us toward unusual transients.  Two of the lowest-mass host galaxies found are SN~2021adxl \citep{Brennan23}, a Hydrogen-rich superluminous SN at $z = 0.018$ and SN\,2021qvo, a 2003fg-like SN\,Ia at $z = 0.05$ (Paniagua et al.\ in prep.).  Both of these SN types tend to favor faint, low-metallicity hosts, and {\tt Blast} reports that both have stellar masses ${\rm log(M_{\ast}/M_{\odot})} \lesssim 8$~dex and metallicities $\lesssim -0.5Z_{\odot}$.  

\begin{figure}
    \centering
    \includegraphics[width=1.0\linewidth]{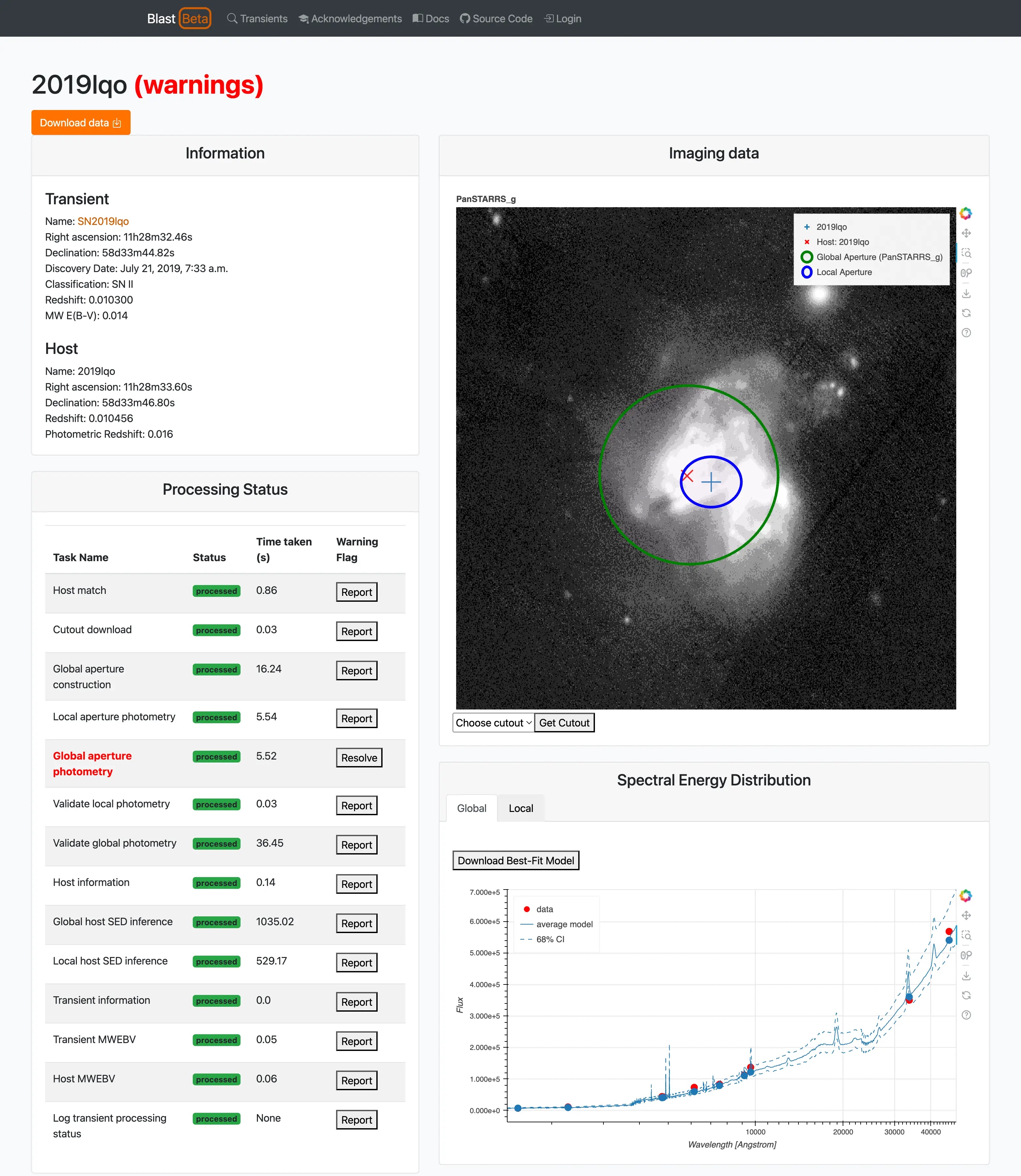}
    \caption{{\tt Blast} page for the merging starburst host of SN~2019lqo, as well as at least seven additional SNe over the last 20~years.  A user has marked the global aperture photometry as spurious due to the difficulty in automated modeling of complex host-galaxy backgrounds.}
    \label{fig:blast_extreme_example}
\end{figure}

On the other hand, SN~2019lqo was hosted by NGC~3690, the most highly star-forming galaxy with mass $>$10~dex in the {\tt Blast} database (Figure \ref{fig:blast_extreme_example}).  NGC~3690 hosted at least six other core-collapse SNe in the last 20~years: SNe 2005U, 2010O, 2010P, 2020fkb, 2022gnp, and 2024gzk, as well as the SN\,Ia 2023wrk.  One of the other examples of a massive, star-forming galaxy was the host of 1991T-like SN~2024gpn, which also hosted the SN\,Ia 2018btm.  One final interesting case is SN~2024crn, which is a SN\,Ia hosted by a galaxy cluster at $z = 0.17$ --- possibly in the intracluster medium --- but coincidentally happened to be 26\arcsec\ from the center of a low-redshift star-forming galaxy, which was spuriously identified as having ${\rm log(M_{\ast}/M_{\odot})} > 12$~dex due to the incorrectly assigned redshift. 

As the {\tt Blast} dataset expands over the coming years, we hope it will provide unprecedented statistical leverage on both rare and common SN subtypes.  We hope that the user community will help to expand its capabilities, and will benefit from {\tt Blast}'s public, transparent SED fitting and host-galaxy photometry for the transient community.

\section{Conclusions}
\label{sec:conclusions}

Here we present {\tt Blast}, a web application for characterizing transient host galaxies.  The goal of {\tt Blast} is to provide transparent, reproducible, public estimation of the physical properties of transient host galaxies and the local regions near where transients occur.  This will help infer the nature of the diverse progenitor systems of astrophysical transients.

{\tt Blast} is constructed in the {\tt Django} framework, implementing a web application and API server that orchestrates a pool of Celery workers to execute tasks including ingesting new transients, downloading archival images, matching transients to a probable host galaxy, computing global and local apertures for photometry, and performing SED-fitting with Prospector as implemented by the SBI$++$ neural posterior estimator \citep{Leja2017,Johnson2021,Wang23}.  All tasks for a given transient are typically completed within 30 minutes after ingestion.  The {\tt Blast} webpages have interactive plots of cutout images and SED fitting results to aid transparency, as well as tables with SED parameters, photometry, and both transient and host-galaxy metadata.  All data are available via the API.

Validation of the different {\tt Blast} stages demonstrates that there are some failure modes at the few-percent level across host-galaxy identification, aperture estimation, photometry, and SED fitting.  While these will continue to be improved in the coming years, we find that {\tt Blast} is largely robust across the diverse set of transient/host-galaxy properties contained in our dataset.  {\tt Blast}'s web-based framework also allows users to easily evaluate whether a given measurement is erroneous and report it.  We report good consistency with other photometric catalogs, and can accurately reproduce independent SED estimation performed in the literature.

We hope to extend {\tt Blast} in the future by incorporating a broader transient redshift range, host galaxy-based transient classifications, and potentially additional SED-fitting algorithms for alternative SPS models or parameter estimation methodologies.  As the Vera Rubin Observatory and the {\tt Roman Space Telescope} come online, and as future LIGO observing runs discover multimessenger objects, there is a novel opportunity for value-added services such as this one to create new scientific opportunities and maximize the potential for discovery in the coming years.


\begin{acknowledgments}

Dedicated to the memory of Tom Jarrett, who was invaluable in helping us to evaluate the success of our SED fitting and work with the WISE data during the preparation of this manuscript.  

D.O.J. and P.M. acknowledge support from NASA grant 80NSSC21K0834. P.M. also acknowledges that this work was partly performed under the auspices of the U.S. Department of Energy by Lawrence Livermore National Laboratory under Contract DE-AC52-07NA27344. The document number is LLNL-JRNL-870752. D.O.J. acknowledges support from HST grants HST-GO-17128.028 and HST-GO-16269.012, awarded by the Space Telescope Science Institute (STScI), which is operated by the Association of Universities for Research in Astronomy, Inc., for NASA, under contract NAS5-26555. D.O.J. also acknowledges support from NSF grant AST-2407632 and NASA grant 80NSSC24M0023.  ST was supported by funding from the European Research Council (ERC) under the European Union's Horizon 2020 research and innovation programmes (grant agreement no. 101018897 CosmicExplorer).  This work is supported by the National Science Foundation under Cooperative Agreement PHY-2019786 (The NSF AI Institute for Artificial Intelligence and Fundamental Interactions, http://iaifi.org/).

Blast’s public interface is operated by the Scalable Cyberinfrastructure for Multi-Messenger Astrophysics (SCiMMA) team under OAC-2311355, supporting the work of TAM and GN, GN also gratefully acknowledges NSF support from NSF CAREER grant AST-2239364, supported in-part by funding from Charles Simonyi, NSF AST 2421845 and support from the Simons Foundation as part of the NSF-Simons SkAI Institute, AST 2432428, AST 2206195, and DOE support through the Department of Physics at the University of Illinois, Urbana-Champaign (\# 13771275), and support from the HST Guest Observer Program through HST-GO-16764. and HST-GO-17128 (PI: R. Foley).

This research uses services or data provided by the Astro Data Lab, which is part of the Community Science and Data Center (CSDC) Program of NSF NOIRLab. NOIRLab is operated by the Association of Universities for Research in Astronomy (AURA), Inc. under a cooperative agreement with the U.S. National Science Foundation.  This research has made use of the NASA/IPAC Infrared Science Archive, which is funded by the National Aeronautics and Space Administration and operated by the California Institute of Technology.

This work used Jetstream2 at Indiana University and Open Storage Network at NCSA through allocation PHY240068 from the Advanced Cyberinfrastructure Coordination Ecosystem: Services \& Support (ACCESS) program, which is supported by National Science Foundation grants \#2138259, \#2138286, \#2138307, \#2137603, and \#2138296.  The SBI$++$ model training was run on the FASRC Cannon cluster supported by the FAS Division of Science Research Computing Group at Harvard University.

The Legacy Surveys consist of three individual and complementary projects: the Dark Energy Camera Legacy Survey (DECaLS; Proposal ID \#2014B-0404; PIs: David Schlegel and Arjun Dey), the Beijing-Arizona Sky Survey (BASS; NOAO Prop. ID \#2015A-0801; PIs: Zhou Xu and Xiaohui Fan), and the Mayall z-band Legacy Survey (MzLS; Prop. ID \#2016A-0453; PI: Arjun Dey). DECaLS, BASS and MzLS together include data obtained, respectively, at the Blanco telescope, Cerro Tololo Inter-American Observatory, NSF’s NOIRLab; the Bok telescope, Steward Observatory, University of Arizona; and the Mayall telescope, Kitt Peak National Observatory, NOIRLab. Pipeline processing and analyses of the data were supported by NOIRLab and the Lawrence Berkeley National Laboratory (LBNL). The Legacy Surveys project is honored to be permitted to conduct astronomical research on Iolkam Du’ag (Kitt Peak), a mountain with particular significance to the Tohono O’odham Nation.

This project used data obtained with the Dark Energy Camera (DECam), which was constructed by the Dark Energy Survey (DES) collaboration. Funding for the DES Projects has been provided by the U.S. Department of Energy, the U.S. National Science Foundation, the Ministry of Science and Education of Spain, the Science and Technology Facilities Council of the United Kingdom, the Higher Education Funding Council for England, the National Center for Supercomputing Applications at the University of Illinois at Urbana-Champaign, the Kavli Institute of Cosmological Physics at the University of Chicago, Center for Cosmology and Astro-Particle Physics at the Ohio State University, the Mitchell Institute for Fundamental Physics and Astronomy at Texas A\&M University, Financiadora de Estudos e Projetos, Fundacao Carlos Chagas Filho de Amparo, Financiadora de Estudos e Projetos, Fundacao Carlos Chagas Filho de Amparo a Pesquisa do Estado do Rio de Janeiro, Conselho Nacional de Desenvolvimento Cientifico e Tecnologico and the Ministerio da Ciencia, Tecnologia e Inovacao, the Deutsche Forschungsgemeinschaft and the Collaborating Institutions in the Dark Energy Survey. The Collaborating Institutions are Argonne National Laboratory, the University of California at Santa Cruz, the University of Cambridge, Centro de Investigaciones Energeticas, Medioambientales y Tecnologicas-Madrid, the University of Chicago, University College London, the DES-Brazil Consortium, the University of Edinburgh, the Eidgenossische Technische Hochschule (ETH) Zurich, Fermi National Accelerator Laboratory, the University of Illinois at Urbana-Champaign, the Institut de Ciencies de l’Espai (IEEC/CSIC), the Institut de Fisica d’Altes Energies, Lawrence Berkeley National Laboratory, the Ludwig Maximilians Universitat Munchen and the associated Excellence Cluster Universe, the University of Michigan, NSF’s NOIRLab, the University of Nottingham, the Ohio State University, the University of Pennsylvania, the University of Portsmouth, SLAC National Accelerator Laboratory, Stanford University, the University of Sussex, and Texas A\&M University.

The Pan-STARRS1 Surveys (PS1) and the PS1 public science archive have been made possible through contributions by the Institute for Astronomy, the University of Hawaii, the Pan-STARRS Project Office, the Max-Planck Society and its participating institutes, the Max Planck Institute for Astronomy, Heidelberg and the Max Planck Institute for Extraterrestrial Physics, Garching, The Johns Hopkins University, Durham University, the University of Edinburgh, the Queen's University Belfast, the Harvard-Smithsonian Center for Astrophysics, the Las Cumbres Observatory Global Telescope Network Incorporated, the National Central University of Taiwan, the Space Telescope Science Institute, the National Aeronautics and Space Administration under Grant No. NNX08AR22G issued through the Planetary Science Division of the NASA Science Mission Directorate, the National Science Foundation Grant No. AST–1238877, the University of Maryland, Eotvos Lorand University (ELTE), the Los Alamos National Laboratory, and the Gordon and Betty Moore Foundation.

Funding for the Sloan Digital Sky Survey IV has been provided by the Alfred P. Sloan Foundation, the U.S. Department of Energy Office of Science, and the Participating Institutions. SDSS acknowledges support and resources from the Center for High-Performance Computing at the University of Utah. The SDSS web site is www.sdss4.org.

SDSS is managed by the Astrophysical Research Consortium for the Participating Institutions of the SDSS Collaboration including the Brazilian Participation Group, the Carnegie Institution for Science, Carnegie Mellon University, Center for Astrophysics | Harvard \& Smithsonian (CfA), the Chilean Participation Group, the French Participation Group, Instituto de Astrofísica de Canarias, The Johns Hopkins University, Kavli Institute for the Physics and Mathematics of the Universe (IPMU) / University of Tokyo, the Korean Participation Group, Lawrence Berkeley National Laboratory, Leibniz Institut für Astrophysik Potsdam (AIP), Max-Planck-Institut für Astronomie (MPIA Heidelberg), Max-Planck-Institut für Astrophysik (MPA Garching), Max-Planck-Institut für Extraterrestrische Physik (MPE), National Astronomical Observatories of China, New Mexico State University, New York University, University of Notre Dame, Observatório Nacional / MCTI, The Ohio State University, Pennsylvania State University, Shanghai Astronomical Observatory, United Kingdom Participation Group, Universidad Nacional Autónoma de México, University of Arizona, University of Colorado Boulder, University of Oxford, University of Portsmouth, University of Utah, University of Virginia, University of Washington, University of Wisconsin, Vanderbilt University, and Yale University.

This publication makes use of data products from the Two Micron All Sky Survey, which is a joint project of the University of Massachusetts and the Infrared Processing and Analysis Center/California Institute of Technology, funded by the National Aeronautics and Space Administration and the National Science Foundation.

This publication makes use of data products from the Wide-field Infrared Survey Explorer, which is a joint project of the University of California, Los Angeles, and the Jet Propulsion Laboratory/California Institute of Technology, funded by the National Aeronautics and Space Administration.
\end{acknowledgments}

\vspace{5mm}

\software{NumPy \citep{Harris20:numpy}, Bokeh \citep{Bokeh18:bokeh}, Astropy \citep{Astropy13:paperI, Astropy18:paperII}, Django \citep{django19:django}, GHOST \citep{Gagliano2021}, MySQL (\url{https://www.mysql.com/}, Docker \citep{merkel2014}, Sphinx \citep{brandl2021}}

\appendix

\section{Prospector-$\alpha$ Priors}
\label{app:priors}

In this section, we list the priors used to generate simulated galaxies for the training of SBI$++$.  Two thirds of our simulated galaxies use the priors in Table \ref{tab:priors_normal}, while the remaining third uses the same parameter ranges but with flat priors across those ranges.  For the 2~kpc local regions, the allowed mass range is 5.5 to 11~dex, while all other priors remain the same.  For additional details on the implementation of the Prospector-$\alpha$ model in SBI$++$, see \citet{Wang23}.

\begin{table}[]
    \centering
    \caption{Priors on the distribution of Prospector-$\alpha$ parameters}
    \begin{tabular}{lrr}
        \hline \\ [-1.5ex]
        Variable$^{\rm a}$ & Prior & Range \\
        \hline \\ [-1.5ex]
        log($M_{\ast}/M_{\odot}$) & flat & 7, 12.5\\
        log($Z_{\ast}/Z_{\odot}$) & \citet{Gallazzi05}$^{\rm b}$ & $-$1.98, 0.19\\
        log($Z_{\rm gas}/Z_{\odot}$) & flat & $-2$, 0.5\\
        SFH & $t(\mu = 0, \sigma=5, \nu=2)^{\rm c}$& $-5$, 5\\
        $\tau_{\rm dust,2}$ & $\mathcal{N}(\mu=0.3,\sigma=1.0)$& 0, 4\\
        $n$ & flat & $-1$, 0.4\\
        $\tau_{\rm dust,1}/\tau_{\rm dust,2}$ & $\mathcal{N}(\mu=1.0,\sigma=0.3)$ & 0, 2\\
        log($f_{\rm agn}$) & flat & $-5$, 0.48\\
        log($\tau_{\rm agn}$) & flat & 0.70, 2.18\\
        $q_{\rm pah}$ & $\mathcal{N}(\mu=2.0,\sigma=2.0)$ & 0, 7\\
        $U_{\rm min}$ &$\mathcal{N}(\mu=1.0,\sigma=10.0)$ & 0.1, 25\\
        ${\rm log}(\gamma_e)$ & $\mathcal{N}(\mu=-2.0,\sigma=1.0)$& $-4$, 0\\
        \hline \\ [-1.5ex]
\multicolumn{3}{l}{
        \begin{minipage}{3.2in}
        $^{\rm a}$ We use the same variables and notation as \citet[their Table 1]{Wang23b}, with the addition of two variables: $U_{\rm min}$, the minimum intensity of stellar radiation that dust grains are exposed to, and log$(\gamma_e)$, the fraction of dust mass exposed to at least this minimum intensity.  All logarithms use base 10.

        $^{\rm b}$ Priors are Gaussian with mean and standard deviation based on the mass-dependent metallicity distribution from  \citet{Gallazzi05}.
        
        $^{\rm c}$ Student's t-distribution.  For details on the adopted age bins, see \citet{Wang23}, their Section 3.3.
        \end{minipage}}
    \end{tabular}
    \label{tab:priors_normal}
\end{table}

\section{Validation of SED Parameters}
\label{sec:app_sed}

\begin{figure*}
    \centering
    \includegraphics[width=7in]{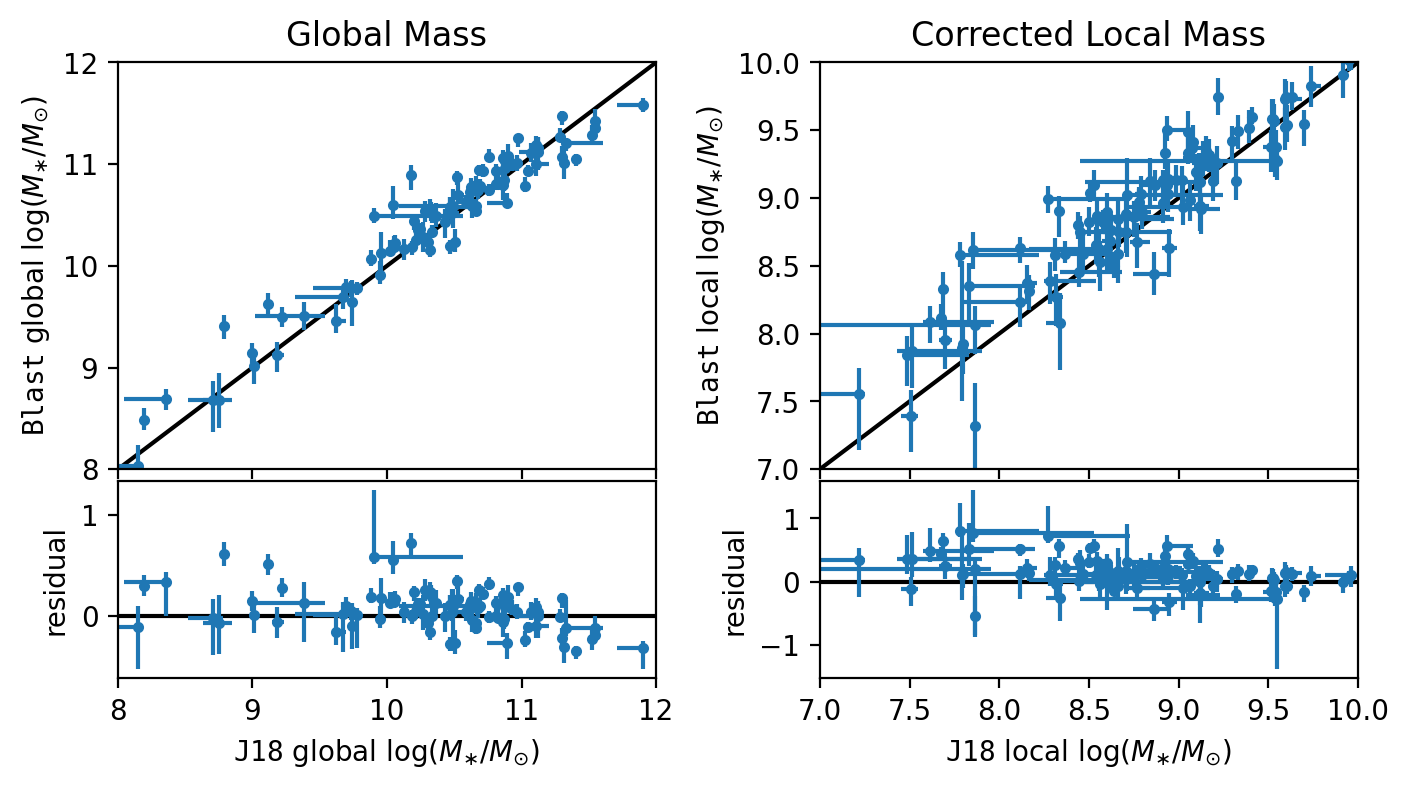}
    \caption{SED parameter comparisons using \citet{Jones2018}-selected SNe for global (left) and local (right) galaxy properties.  Local properties from {\tt Blast} are reduced by 0.25~dex due to the approximate difference in aperture area between this work and J18 (2~kpc versus 1.5~kpc), which is likely responsible for the systematic offsets at the low-mass end.}
    \label{fig:sedcomp_j18}
\end{figure*}

\begin{figure}
    \centering
    \includegraphics[width=3.4in]{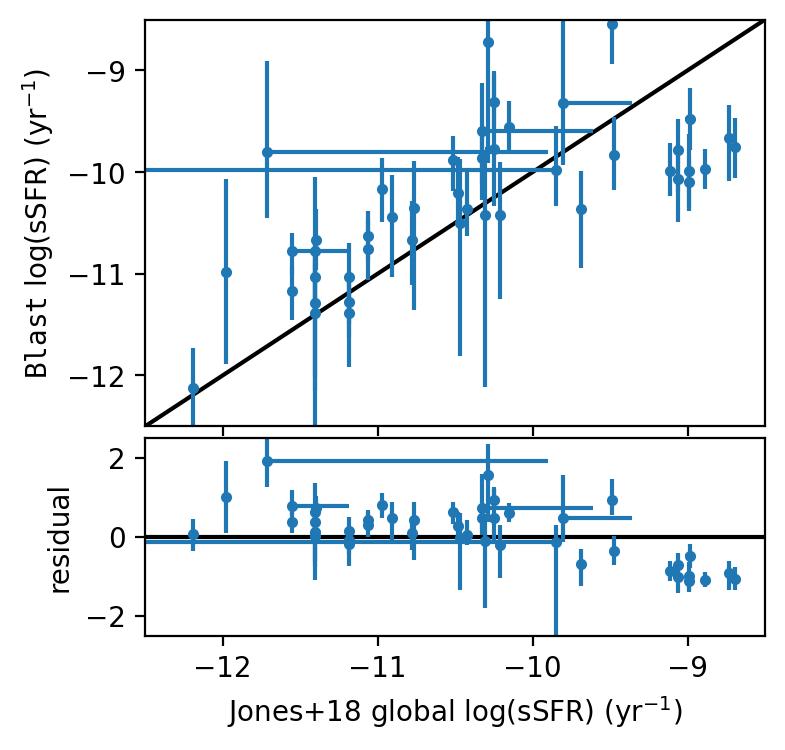}
    \caption{Global sSFR parameter comparisons between {\tt Blast} and LePHARE $ugrizy$-derived quantities from \citet{Jones2018}.  Agreement is good overall, but {\tt Blast} prefers lower sSFR for the most extreme star-forming objects in the sample.}
    \label{fig:ssfrcomp_j18}
\end{figure}

\begin{figure*}
    \centering
    \includegraphics[width=7in]{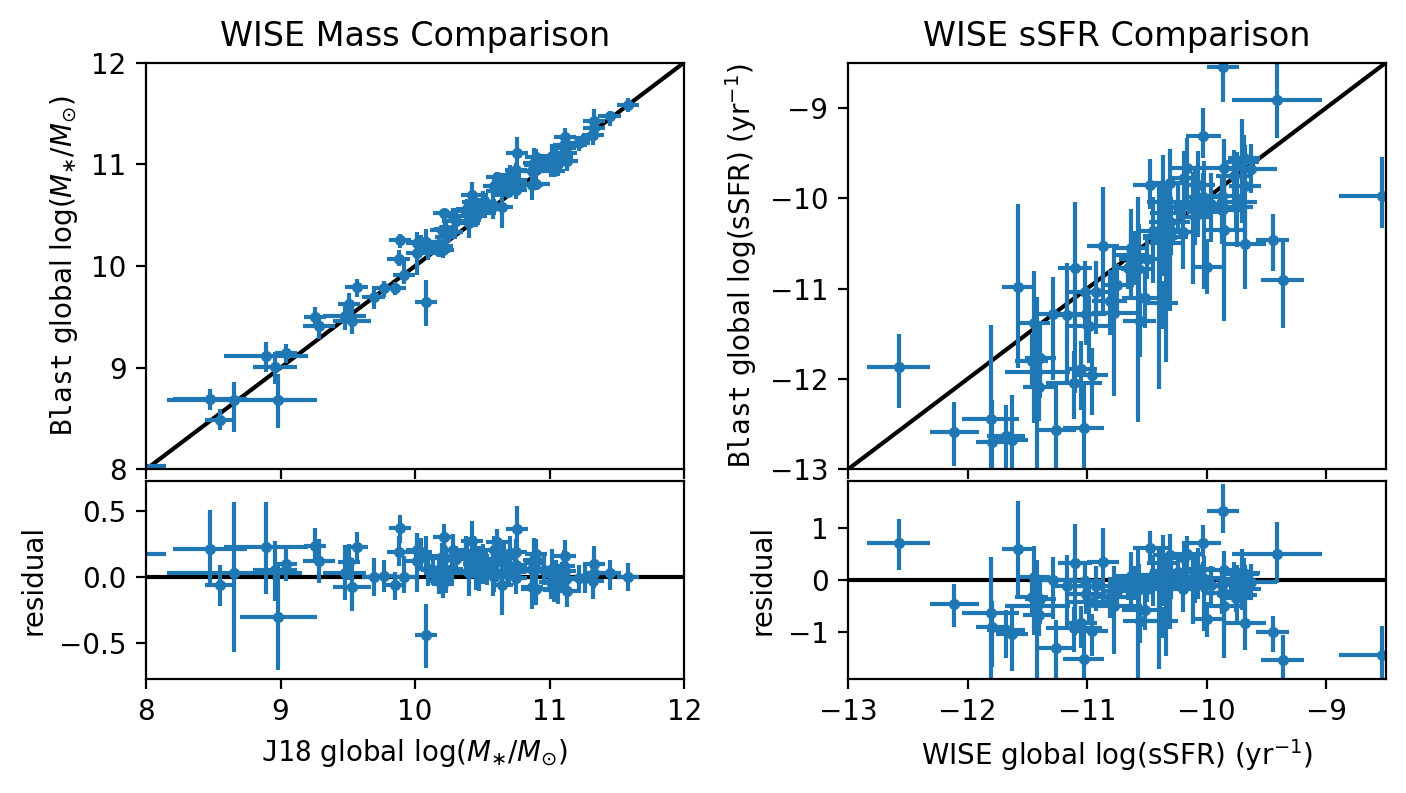}
    \caption{Comparison between {\tt Blast} and quantities derived from the WISE infrared bands following \citet{Jarrett23}.  Mass agrees closely (a median difference of just 0.06~dex), while {\tt Blast} reports slightly lower sSFR by 0.38~dex on average.}
    \label{fig:sedcomp_jarett}
\end{figure*}

\begin{figure*}
    \centering
    \includegraphics[width=7in]{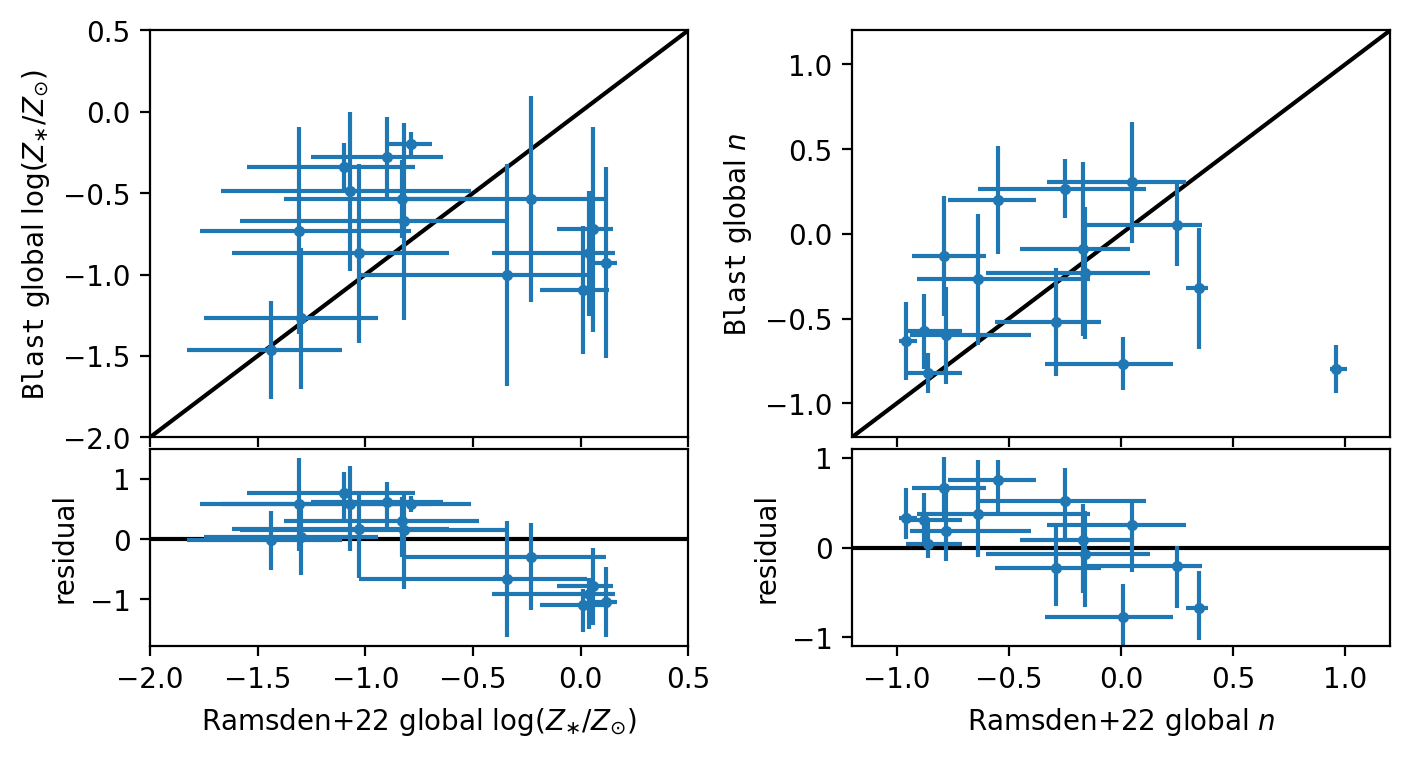}
    \caption{Comparison between \citet{Ramsden22} and {\tt Blast}, for two of the more uncertain Prospector-$\alpha$ parameters: the stellar metallicity, ${\rm log}_{10}(Z_{\ast}/Z_{\odot})$, and the dust power-law index, $n$.  Agreeement is typically within the errors, and there is a median offset of 0.05~dex for stellar metallicity and $\sim$0.1 for the dust optical depth.  Inspection of individual outliers shows slightly different photometry and photometric bands included in the fit.}
    \label{fig:ramsden}
\end{figure*}

This appendix contains several figures comparing {\tt Blast}'s SED-fitting parameter estimation to those from other publications and methodologies.  Figure \ref{fig:sedcomp_j18} shows stellar mass, Figure \ref{fig:ssfrcomp_j18} shows sSFR, Figure \ref{fig:sedcomp_jarett} shows stellar mass and sSFR compared to inferences from WISE IR data, and Figure \ref{fig:ramsden} shows comparisons to stellar metallicity and dust power-law index using the small sample from \citet{Ramsden22}.

\clearpage
\bibliography{sample631}{}
\bibliographystyle{aasjournal}



\end{document}